%% file: core.tex
\documentclass[journal]{IEEEtran}

\usepackage{amsmath,amssymb,amsfonts}
\usepackage{array} 
\newcolumntype{P}[1]{>{\centering\arraybackslash}p{#1}}
\newcolumntype{M}[1]{>{\centering\arraybackslash}m{#1}}
\usepackage{adjustbox}
\usepackage{epstopdf}
\usepackage{multirow}
\usepackage{subfiles}
\usepackage{pgfplots}
\usepackage{pgfplotstable}

\pgfplotsset{compat=1.7}
\usepackage{pdfpages}
\usepackage[super]{nth}
\usepackage{tikz}

\usepackage{hyperref}
\usepackage{afterpage}
\usepackage{ulem}

\newcommand*\circled[1]{\tikz[baseline=(char.base)]{
            \node[shape=circle,draw,inner sep=2pt] (char) {#1};}}

\usetikzlibrary{fit,positioning}
\tikzstyle{every picture}+=[remember picture]

\input{sections/commands}

\ifCLASSOPTIONcompsoc
\usepackage[caption=false,font=normalsize,labelfon
t=sf,textfont=sf]{subfig}
\else
\usepackage[caption=false,font=footnotesize]{subfi
g}
\fi

\begin{document}

\title{Data-Driven Interpolation for Super-Scarce \xr Computed Tomography}

\author{Emilien~Valat,
        Katayoun~Farrahi,
        and~Thomas~Blumensath
\thanks{Emilien V., Katayoun F. and Thomas B. are with the University of Southampton, Southampton, United Kingdom}
\thanks{This work is supported by the DSTL/DGA PhD scheme.}
}
\markboth{}
{Shell \MakeLowercase{\textit{et al.}}: Data-Driven Interpolation for Limited Angles \xr Computed Tomography}
\maketitle

\begin{abstract}
    We address the problem of reconstructing X-Ray tomographic images from scarce measurements by interpolating missing acquisitions using a self-supervised approach. To do so, we train shallow neural networks to combine two neighbouring acquisitions into an estimated measurement at an intermediate angle. This procedure yields an enhanced sequence of measurements that can be reconstructed using standard methods, or further enhanced using regularisation approaches. 

    Unlike methods that improve the sequence of acquisitions using an initial deterministic interpolation followed by machine-learning enhancement, we focus on inferring one measurement at once. This allows the method to scale to 3D, the computation to be faster and crucially, the interpolation to be significantly better than the current methods, when they exist. We also establish that a sequence of measurements must be processed as such, rather than as an image or a volume. We do so by comparing interpolation and up-sampling methods, and find that the latter significantly under-perform.
    
    We compare the performance of the proposed method against deterministic interpolation and up-sampling procedures and find that it outperforms them, even when used jointly with a state-of-the-art projection-data enhancement approach using machine-learning. These results are obtained for 2D and 3D imaging, on large biomedical datasets, in both projection space and image space. 

\end{abstract}


\IEEEpeerreviewmaketitle

\section{Introduction}

\subfile{sections/introduction}

\section{Contributions}

\subfile{sections/contribution}

\section{Proposed Approach}

\subfile{sections/proposedApproach}

\section{Implementation Details}

\subfile{sections/implementationDetails}

\section{Datasets}

\subfile{sections/Datasets}

\section{Results}

\subfile{sections/results}

\section{Discussion and Conclusion}

\subfile{sections/conclusion}

\bibliographystyle{ieeetr}
\bibliography{sections/bibliography.bib}

\section{Images}

\subfile{sections/images}

\section{Tables}

\subfile{sections/tables}

\section{Figures}

\subfile{sections/figures}

\end{document}

%% file: sections/commands.tex
\newcommand{\xr}{X-Ray }

\makeatletter
\newcommand{\thickhline}{%
    \noalign {\ifnum 0=`}\fi \hrule height 1pt
    \futurelet \reserved@a \@xhline
}
\newcolumntype{"}{@{\hskip\tabcolsep\vrule width 1pt\hskip\tabcolsep}}
\makeatother

%% file: sections/introduction.tex
\IEEEPARstart{X}Ray Computed Tomography (XCT) is a non-destructive imaging technique that reconstructs cross-sectional images of objects whose materials interact with this radiation. XCT uses projection images acquired from different viewpoints around the object, and final image quality relies on the number of acquisitions as well as their uniform distribution around the object.  As such, there is often a trade-off between the performance of the imaging modality and physical constraints such as X-ray toxicity and imaging time. Sparse-data sampling is a common mitigating strategy that speeds up acquisition and limits  \xr exposure by reducing the number of acquisitions. Advanced algorithms are then required to effectively streamline the image reconstruction process. 

\begin{figure}
\centering
\begin{tikzpicture}[
    node distance = 7mm and -3mm,
every node/.style = {draw=black, rounded corners, fill=gray!40, 
                     minimum width=0cm, minimum height=0.5cm,
                     align=center}
                        ]
\node (0) {Object};
\node (1)[below=5mm of 0.south] {Projections};
\node (2)[below=5mm of 1.south] {Enhanced \\ Projections};
\node (3)[below=5mm of 2.south] {Image};
\node (4)[below=5mm of 3.south] {Enhanced \\ Image};

\node (5) [below right=2.5mm and 10mm of 0.east, fill=gray!10] {\scriptsize{Sampling Process.}};

\node (6) [below=5mm of 5.south, fill=gray!10] {\scriptsize{Projection data} \\ \scriptsize{Enhancement.}};

\node (7) [below=7.5mm of 6.south, fill=gray!10] {\scriptsize{Image Reconstruction.}};

\node (8) [below=5mm of 7.south, fill=gray!10] {\scriptsize{Image Improvement.}};

\node (9) [left=7.5mm of 2.west, fill=gray!10] {\scriptsize{Image}\\ \scriptsize{Reconstruction.}};

\draw [->] (0.south) to (1.north) ;
\draw [->,blue] (1.south) to (2.north);
\draw [->,green] (2.south) to (3.north);
\draw [->,red] (3.south) to (4.north);

\draw [->] (1.west) to [out=220,in=180] (3.west);

\end{tikzpicture}
\caption{Data-driven regularisation frameworks. The blue line represents a \textit{projection data enhancement} method, whilst the green (resp. red) line represents an \textit{image reconstruction} (resp. \textit{image improvement}) method. The black lines represent the minimal transformations (sampling, reconstruction) necessary to produce the cross-sectional volume. The Light-gray nodes label the transformations whilst the dark-gray nodes represent the different states of the data throughout the imaging process.}
\label{figure:XCT_improvement}
\end{figure}
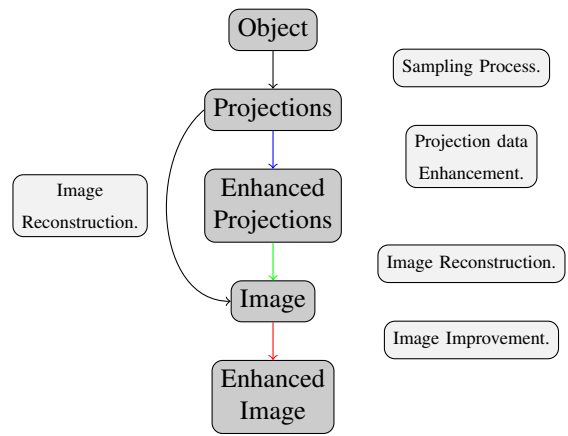

Those procedures are called regularisation frameworks, and they are various. Traditional methods, such as total variation regularised reconstruction, try to reconstruct smooth images with sharp edges whilst more recent approaches often take a data driven approach, where problem specific regularisation strategies are learned from larger training data sets using modern machine learning techniques. Data-driven regularisation can either be included into an \textit{image reconstruction}, an \textit{image improvement} or a \textit{projection data enhancement} framework. These three can be combined, as shown in Fig. \ref{figure:XCT_improvement}.

In this paper, we focus only on the latter framework, which consists of the estimation and de-noising of missing acquisitions. The sequence of acquisitions is typically stacked into 2D or 3D images (so called sinograms) and several methods use standard de-noising and interpolation tools directly on these sinograms. For instance, \cite{Yuan2018} uses an initial bicubic interpolation to estimate missing data in the sinogram, which is then further improved using a U-net \cite{Ronneberger2015} deep neural network, whilst \cite{Lee2018} uses a linear interpolation followed by a U-net. \cite{Pelt2022} also uses bicubic spline interpolation and convolutional neural network enhancement, whilst  \cite{Ernst2021} uses bilinear interpolation followed by a regularisation performed during reconstruction, using the regularised primal-dual algorithm \cite{Adler2017}. The common theme here is that standard interpolation methods are used followed by learned enhancement algorithms. This limits the performance in the following ways:
\begin{itemize}
	\item Networks capable of processing the entire sinogram require significant computer memory, and are slower to train.
	\item Training requires many sinograms which are not normally available.
	\item The initial (deterministic) interpolation performs relatively poorly on sinogram data.
\end{itemize}

We thus propose the use of a data-driven interpolation step, that optimally interpolates the two closest neighbouring projections images. First, this approach requires significantly less memory for each computation, as only two acquisitions are processed at once instead of the whole sinogram (or patches of it). As a consequence, it is parallelisable by design. Indeed, instead of training one large model for all combinations and distribute it across several Graphical Processing Units (GPU), we train several shallow models on different GPU, but without distribution of their parameters. Then, this is more efficient with the available training data, as a single set of projection provides many training samples. Finally, it provides a more optimal initial interpolation than deterministic interpolation methods enhanced with image processing tools, and can be further combined with the latter. Crucially, our approach is scalable, and thus applicable to realistic 3D data-sets.

%% file: sections/contribution.tex
This paper proposes the learnt interpolation operation as a framework for reducing X-Ray intake in XCT. This approach relies convolutional neural networks, trained in a self-supervised fashion by minimising the mean-squared error of the infered acquisitions.

Using that approach, we show that interpolating acquisitions is a better choice than up-sampling a sinogram, using either deterministic or data-driven techniques. Crucially, we prove that doing a poor initial guess and enhancing it using state-of-the-art image improvement techniques is a worst choice than just carefully interpolating the acquisitions. For 6.25\% of available acquisitions, the average improvement gain against other interpolation methods is superior to 6 dB PSNR, in projection and image space, on two biomedical datasets for 2D and 3D XCT.

By experimenting on four different proportions of available acquisitions, we show that it is better to use a 12.5\% of acquisitions and interpolate the rest using our method, rather than 25\% and interpolating the measurements with deterministic methods. These results are obtained for 3D XCT, in both projection and image space.

To have these results, we designed a computational biomedical dataset for this project, which is available for the community to experiment their own algorithms with. Additionally, our code and the learned network parameters are available for reproducibility.

%% file: sections/proposedApproach.tex
We study a tomographic imaging setup in which limited measurements are made at equal intervals around an object, and design bespoke neural networks to predict the missing acquisitions. The aim is to design an efficient and fast interpolation strategy that is easily scalable to large 3D datasets. 

\subsection{The Interpolation Problem}
Given a scarce-view sinogram made of N acquisitions sampled at regular angular intervals $\Delta\theta$, our goal is to infer intermediate measurements at given angular positions to up-sample the sinogram by a chosen ratio R. Let $A_i$ be a sampled (reference) measurement, a linear interpolation of acquisitions at equally distant intervals would compute the following:
    
\begin{equation}
A_j = w A_i + (1-w) A_{i+\Delta\theta}  
\label{eq:linear_interpolation}
\end{equation}
With
\begin{equation*}
    w = 1-\frac{j-i}{\Delta\theta}, \forall j \in \left[ i+1, i+\mathrm{R}-1 \right] 
\end{equation*}

Other interpolation methods estimate the contribution of neighbouring acquisitions and then use a weighted sum of these estimates to predict the intermediate acquisition. Consider Eq. (\ref{eq:linear_interpolation}), and let $g_w$ be a function that maps $A_i$ to $A_j$:
\begin{equation}
\label{superSlowMo_interpolation_equation}
A_j = w g_w(A_i) + (1-w) g_{1-w}(A_{i+\Delta\theta})
\end{equation}
The approach described in Eq. (\ref{superSlowMo_interpolation_equation}) is used in state-of-the-art (SOTA) interpolation techniques for estimating missing frames in videos, such as SuperSlowMo \cite{Jiang2017}. More generally, one can write the following generic interpolation:
\begin{equation}
\label{our_computation}
A_j = f_w (A_i , A_{i+\Delta\theta})
\end{equation}
Where $f_w$ is a non-linear function depending on R and $\Delta\theta$. We propose to approximate $f_w$ using a neural network (NN). 

\subsection{Design of the Interpolation Networks}
\label{subsection:design_of_interpolation_networks}

\begin{figure}
\centering
\begin{tikzpicture}[
    node distance = 7mm and -3mm,
every node/.style = {draw=black, rounded corners, fill=gray!30, 
                        minimum width=0cm, minimum height=0.5cm,
                        align=center}
                        ]
\node (0) {$A_0$};
\node (1)[right=5mm of 0.east] {$A_1$};
\node (2)[right=5mm of 1.east] {$A_2$};
\node (3)[right=5mm of 2.east] {$A_3$};
\node (4)[right=5mm of 3.east] {$A_4$};

\draw [->,red] (4.north) to [out=150,in=30] (1.north);
\draw [->,green] (4.north) to [out=150,in=30] (2.north);
\draw [->,blue] (4.north) to [out=150,in=30] (3.north);

\draw [->,blue] (0.south) to [out=-30,in=-150] (1.south);
\draw [->,green](0.south) to [out=-30,in=-150] (2.south);
\draw [->,red]  (0.south) to [out=-30,in=-150] (3.south);

\end{tikzpicture}
\caption{For a factor of reduction of acquisitions of 4, there are 2 ways to combine evenly spaced acquisitions in between the two reference ones. }
\label{fig:combining_acquisitions}

\end{figure}
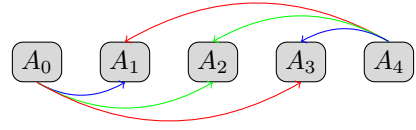

As one can observe in Fig. \ref{fig:combining_acquisitions}, acquisitions $A_0$ and $A_4$ are given and we would like to infer missing ones, at three intermediate angles. Using Eq. \ref{eq:linear_interpolation}, we have

\begin{equation}
\begin{split}
A_1 &= \frac{3}{4} A_0 + \frac{1}{4} A_4 = f_{(\frac{1}{4}, \frac{3}{4})} (A_0, A_4)\\
A_2 &= \frac{2}{4} A_0 + \frac{2}{4} A_4 = f_{(\frac{2}{4}, \frac{2}{4})} (A_0, A_4)\\
A_3 &= \frac{3}{4} A_4 + \frac{1}{4} A_0 = f_{(\frac{1}{4}, \frac{3}{4})} (A_4, A_0)
\end{split}
\end{equation}
We train two networks to map $f_{(\frac{1}{4}, \frac{3}{4})}$ and $f_{(\frac{2}{4}, \frac{2}{4})}$. This choice imposes two constraints on the NN design: it must have two input channels on the first layer and a network will always receive the input from the closest acquisition on the first channel of its input layer.

%% file: sections/implementationDetails.tex
We our proposed approach, based on the U-net architecture, and the image-improvement process of \cite{Lee2018} for 2D XCT.

\subsection{U-nets Architecture for Combining Acquisitions}
\label{unets_architecture}
According to the constraint stated in \ref{subsection:design_of_interpolation_networks}, the architecture has a two-channel input layer. For 2D (resp. 3D) imaging, layers have 1D (resp. 2D) kernels. 

Let $C(n, m, k, s, p)$ be a convolutional layer with  $n$ input features, $m$ output features, a kernel of size $k$, a stride of size $s=1$ and a padding of size $p=1$ (default), followed by a LeakyReLu activation function with a negative slope of $0.1$. Let \textbf{$D_{m,n}$} be down-sampling unit with $m$ input features and $n$ output features. It is composed of the following sequence of layers
\begin{center}
	$C(m, m, 4, 2, 1$) - $C(m, n, 3)$ - $C(n, n, 3)$
\end{center}
The encoding part of the U-net is implemented as follows:
\textbf{$C(2,32,3)$} - \textbf{$C(32,32,3)$} - \textbf{$D_{32-64}$} - \textbf{$D_{64-128}$} - \textbf{$D_{128-256}$} - 
\textbf{$D_{256-512}$} - 
\textbf{$D_{512-1024}$} - 
\textbf{$D_{1024-1024}$} - 
\textbf{$D_{1024-1024}$} - 

Let $CT(m, n, k, s, p)$ be a transposed convolutional layer with  $m$ input features, $n$ output features, a kernel of size $k$, a stride of size $s=1$ and a padding of size $p=1$ (default), followed by a LeakyReLu activation function with a negative slope of $0.1$. Let \textbf{$U_{m,n}$} be an up-sampling unit with $m$ input features and $n$ output features. It is composed of the following sequence of layers:
\begin{center}
	$CT(m, m, 4, 2, 1$) - $C(m, n, 3)$ - $C(2n, n, 3)$
\end{center}
Note that the last layer of \textbf{$U_{m,n}$} has two $2n$ input features has it receives the residual from the corresponding down-sampling unit. The decoding part of the U-net is implemented as follows:
\textbf{$D_{1024-1024}$} - 
\textbf{$D_{1024-1024}$} - \textbf{$D_{1024-512}$} - 
\textbf{$D_{512-256}$} - 
\textbf{$D_{256-128}$} - 
\textbf{$D_{128-64}$} - 
\textbf{$D_{64-32}$} - 
\textbf{$C(32,2,3)$} - \textbf{$C(4,1,3)$}

\subsection{Sinogram Improvement Implementation}
\label{sinogram_improvement_implementation}
To study the impact of the initial interpolation on further regularisation frameworks, we implement the 2D sinogram improvement method proposed in \cite{Lee2018} and compare the performance it yields on sinograms interpolated with various methods. We do not implement it for 3D XCT as it does not scale to 3D.
We implement this patch-based method, though use patches of size 64x64 for computational convenience instead of 50x50, which according to them should not affect performances. As patches are processed, all convolutional layers are 2D. The encoding part of the U-net is implemented as follows:
\textbf{$C(1,32,3)$} - \textbf{$C(32,32,3)$} - \textbf{$D_{32-64}$} - \textbf{$D_{64-128}$} - \textbf{$D_{128-256}$} - 
\textbf{$D_{256-512}$} - 
\textbf{$D_{512-1024}$}
The decoding part of the U-net is implemented as follows:
\textbf{$D_{1024-512}$} - 
\textbf{$D_{512-256}$} - 
\textbf{$D_{256-128}$} - 
\textbf{$D_{128-64}$} - 
\textbf{$D_{64-32}$} - 
\textbf{$C(32,2,3)$} - \textbf{$C(4,1,3)$}

\subsection{Training}
We use the Adam optimiser with a learning rate of $10^{-4}$, train each network for four epochs by minimising the Mean-Square-Error between the inferred and target acquisitions. We refer the reader to our \href{https://github.com/Emvlt/data-driven-interpolation}{GitHub} and our \href{https://tensorboard.dev/experiment/WJEZyqVDSBe3sNmI350jAg}{Tensorboard} repositories for further implementation informations.

%% file: sections/datasets.tex
We compare the methods using two datasets that cover various applications in medical imaging and industrial testing. Re-projected data and projected data are considered for both 2D and 3D cases. The sampling and reconstruction of the CT data uses the TIGRE \cite{Biguri2016, Biguri2020} tomography toolbox and Table 1 shows the different scanning geometries used. For a fast reconstruction, with use the Simultaneous Iterative Reconstruction Technique algorithm with 10 iterations, we use the SART \cite{Andersen1984} algorithm with 3 iterations for reconstructions used for display. This choice is made to have a quick computation of the reconstructions for reporting the results, and a longer but better one for the display. All the sinograms are normalised after sampling.

\subsection{The Lung CT Dataset}
This dataset \cite{Grove2015} consists of reconstructed lung scans from 60 patients for a total of 4682 slices. Out of these scans, we chose to exclude the scans of the patients 232 and 233 due to an uncertainty on the file format that is beyond of our scope of understanding. As a result, we use slices from 58 patients, that we split into 53 for training and 5 for testing, for a total of 4543 slices. Each slice is then re-projected for 2D tomography: using the scanning geometry shown in Table \ref{table:scanning_geometries}, the dataset consists of 2330559 1D acquisitions. Fig. \ref{figure:R_111_patient_slice} shows a slice from the dataset.

\subsection{The 3D Hand Model Dataset}
This dataset consists of 50 3D left-hand models. It consists of a rigged blender model that can be deformed in order to change various parameters affecting the pose of the hand and the shape of fingers. As such, the dataset is a single blender file with a random deformation script. For reproducibility purposes, we indicate that we use the seed 22032022 to generate the dataset and rely on numpy random library. The dataset is available at \cite{Valat2022Hands}. The models are point clouds representing the surface of the object: no intern features are simulated. The base model and the code will be made available shortly after the publication of the paper.
For this study, we split the dataset into 45 hands for training and 5 hands for testing. Fig. \ref{figure:3D_hand_example} shows the baseline model from which the samples are derived, using random transformations of the bone rig. For instance, Fig. \ref{figure:sample_7} shows the \nth{7} sample of the dataset from three different viewpoints. When re-projected using the scanning geometry shown in Table \ref{table:scanning_geometries}, the dataset consists of 25650 2D acquisitions.

%% file: sections/results.tex
Our goal is to assess the efficiency of our interpolation method for sinogram enhancement, its impact on further regularisation techniques, scalability to 3D XCT, and efficiency against the spacing between acquisitions. As such, we compare it to other interpolation methods by assessing the quantitative improvement yielded by the various methods on the sinogram and the reconstructed image. Experiments are performed for an up-sampling ratio R of 16, that is, in the setup described in \ref{table:scanning_geometries}, interpolating a 33-acquisition sinogram to a 513 one.

\subsection{Description of Experiments}
To begin with, for 2D XCT, we compare the performances of three deterministic methods (\textit{linear interpolation}, \textit{bilinear up-sampling}, \textit{nearest neighbours up-sampling}) to ours. We also estimate the impact of 2D enhancement procedures by implementing and training \cite{Lee2018} on sinograms interpolated with the four methods mentioned above. Then, for 3D XCT, we compare \textit{trilinear up-sampling}, \textit{linear interpolation} to our proposed approach. Finally, we change the proportion of given acquisitions to train our method and assess its performance for various R.

We measure performance using the Peak Signal-to-Noise Ratio and compare the estimated sinograms to the ground truth (Sinogram-PSNR), as well as comparing the reconstructed images computed from the estimated sinograms using the SIRT algorithm (Image-PSNR). For sinogram estimation comparison, we further list the performance as a function of angular distance (Angular-PSNR). PSNR between arrays A and B is computed as :
\begin{equation*}\label{eqn:psnr}
    \mbox{PSNR(A,B)} = 20 \log_{10}(\mbox{max(A,B)}) - 10\log_{10}(\mbox{MSE(A,B)})
  \end{equation*}
With MSE(A,B) the Mean-Square-Error between A and B. 

Sinogram-PSNR provides a general view of the quality of each method, whilst angular-PSNR allows us to understand the relationship between interpolation method and angular distance. Finally, image-PSNR shows if meaningful information in the image domain is added by the interpolation.

\subsection{2D LungCT Results}
\paragraph{Comparison between Interpolation Methods}
The results are assessed on the test dataset, which is made of the slices from the patients with ID 64, 127, 143, 146 and 274. To begin with, we assess the sinogram-PSNR of sinograms interpolated using different up-sampling and interpolation methods and present the results in Table \ref{table:2D_LungCT_sinogram-PSNR}. For all volumes considered, we observe that the proposed approach yields an improvement of the sinogram-PSNR of 5.4db PSNR compared to the linear interpolation method, when averaged across the whole test dataset. Up-sampling methods significantly under-perform compared to interpolation methods. In addition to that, we report the per-slice sinogram-PSNR for the sample 274 in Fig. \ref{figure:sinogram_PSNR_274}. For all slices considered, the proposed approach outperforms the other methods. Fig. \ref{figure:sinograms_274_60} shows, for the slice 6 of the sample 274 of the HandCT dataset, the sinograms interpolated using the four described methods and their differences with the target sinogram.

Considering the angular-PSNR shown in Fig. \ref{figure:angular_PSNR_LungCT}, it is clear that the initial enhancement method influences the sinogram quality. For the nearest-neighbour up-sampling, the performance drops significantly for the acquisitions further away from the reference ones. For bilinear up-sampling, the drop is less important and it averages to a better sinogram-PSNR. The proposed method outperforms deterministic methods for all considered acquisitions.

This performance is confirmed by the image-PSNR, as shown in Table \ref{table:2D_LungCT_reconstruction-PSNR}. The proposed approach yields a 6.1dB PSNR improvement compared to linear interpolation when averaged across the whole test dataset. We also report the per-slice image-PSNR for the sample 274 in Fig. \ref{figure:image_PSNR_274} and the results are consistent with the per-slice sinogram-PSNR. A qualitative comparison between the target and scarce reconstructions, and the ones obtained from the linear and proposed interpolation method is available in Fig. \ref{figure:LungCT_reconstructions}.

\paragraph{Impact on Further Regularisation}
We report the impact of the initial interpolation on the 2D sinogram enhancement procedure in Table \ref{table:2D_LungCT_sinogram-PSNR_improved}. The procedure enhances significantly the sinogram-PSNR of the deterministic sinogram interpolation and up-sampling methods. However, the performance gain is less important for the proposed approach. As detailed in Table \ref{table:2D_LungCT_sinogram-PSNR_improved}, the 2D enhancement procedure yields only a 0.8 dB PSNR. Crucially, up-sampling a sinogram and enhancing it using an image-processing tool does not outperform the proposed interpolation method. This results is confirmed by the angular-PSNR shown in Fig. \ref{figure:angular_PSNR_LungCT_improved} and the image PSNR shown in Fig. \ref{figure:image_PSNR_LungCT_improved}. The whole 2D enhancement procedure yields a 0.72dB PSNR improvement, compared to the proposed stand-alone interpolation approach. 

\subsection{3D HandCT Results} 
\paragraph{Comparison Between Interpolation Methods}
The results are assessed on the test dataset, which is made of the volumes with ID 7, 18, 3, 2 and 46. To begin with, we assess the sinogram-PSNR of the sinograms interpolated using different up-sampling and interpolation methods and present the results in Table \ref{table:3D_handCT_sinogram-PSNR}. 

For all the volumes considered, we observe that the proposed approach yields an improvement of the sinogram-PSNR of 6.48dB PSNR compared to the linear interpolation method, when averaged accross the test dataset. Compared to trilinear up-sampling, the proposed method yields an average sinogram-PSNR enhancement of 7.86 dB. The angular-PSNR shown in Fig. \ref{figure:angular_PSNR_HandCT} shows that up to 9.14dB are gained on acquisitions inferred at an angular distance of 8. compared to linear interpolation.

This performance is confirmed by the image-PSNR, as shown in Table \ref{table:3D_handCT_reconstruction-PSNR}: the proposed approach yields a 7.4dB PSNR improvement compared to linear interpolation when averaged accross the test dataset, and 10.56dB PSNR compared to trilinear up-sampling. Fig. \ref{figure:acquisitions_HandCT} shows acquisitions inferred using the three considered methods, and their difference to the target acquisition. 

\paragraph{Efficiency Against Up-sampling Ratio}
We reproduce the whole enhancement procedure for up-sampling ratios of 2, 4 and 8 to assess the performance of our method for different amount of acquisitions. These ratios correspond to angular intervals of 0.7\textdegree, 1.4\textdegree and 2.8\textdegree between available acquisitions. The results for the sinogram-PSNR averaged accross the test dataset are shown in Fig. \ref{figure:Sinogram-PSNR_against_ratio}. It is clear that, for an up-sampling ratio of 2, linear interpolation performs well. However, for fewer available projections, our method outperforms deterministic interpolation and up-sampling methods. The performance on the sinogram must be mitigated by the results of the image-PSNR shown in Fig. \ref{figure:Image-PSNR_against_ratio}. In the image space, the improvement is noticeable only for up-sampling ratios of 8 and 16.

%% file: sections/conclusion.tex
This paper presents the first self-supervised method for interpolating missing acquisitions in XCT sinograms. To begin with, we show that up-sampling a sinogram was an underperforming choice and that interpolating acquisitions yields an immediate and significant improvement of the sinogram quality. We then demonstrate that training a NN to combine acquisitions based on their angular distances was an efficient way to infer missing measurements in the sinogram. After, we show that our method could scale to 3D XCT by assessing its performance on a large 3D dataset. Finally, we show that method works for different proportions of available projection data, but that the improvement yielded was significant for 12.5\% of acquisitions and below.

%% file: sections/images.tex
\begin{figure}
{\includegraphics[width=0.4\textwidth]{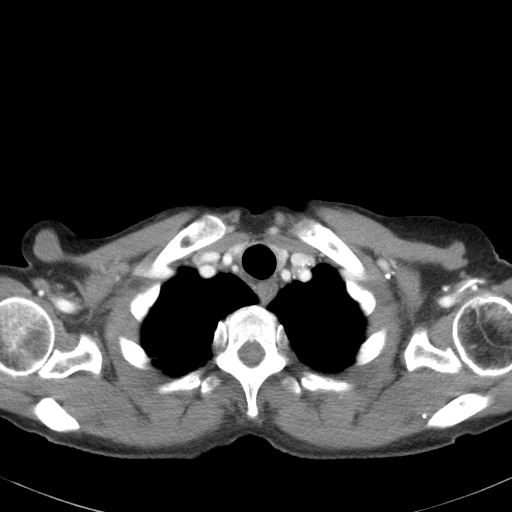}}
\caption{Slice from the patient 111 from the Lung CT dataset.}
\label{figure:R_111_patient_slice}
\end{figure}

\begin{figure}
{\includegraphics[width=0.4\textwidth]{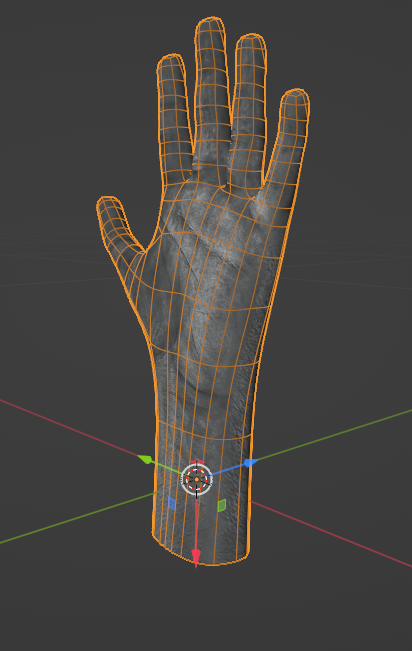}}
\caption{3D hand sample with default settings. To generate a sample, the hand is deformed using a script that applies random changes to the rig of the model.}
\label{figure:3D_hand_example}
\end{figure}

\begin{figure*}

\centering
\subfloat[0 \textdegree view]{\includegraphics[width=0.2\textwidth]{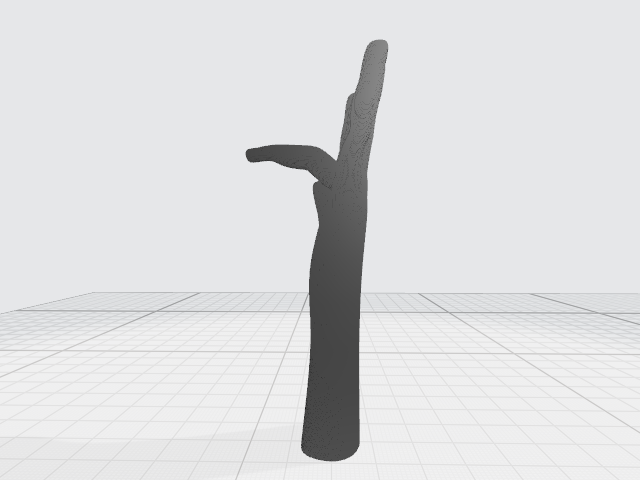}}
\label{figure:hand_7_0}
\subfloat[45 \textdegree view]{\includegraphics[width=0.2\textwidth]{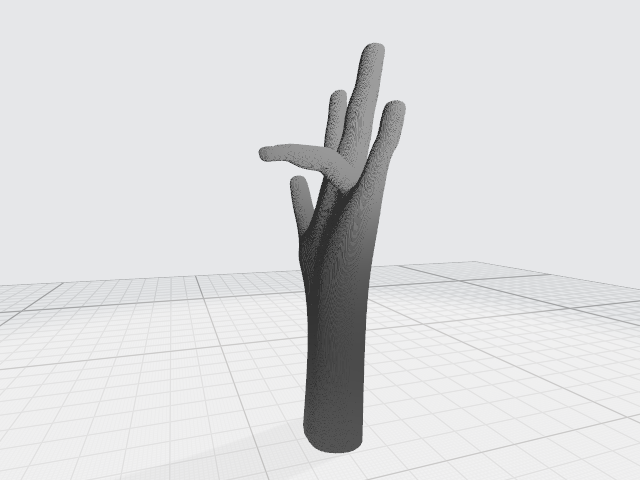}}
\label{figure:hand_7_45}
\subfloat[90 \textdegree view]{\includegraphics[width=0.2\textwidth]{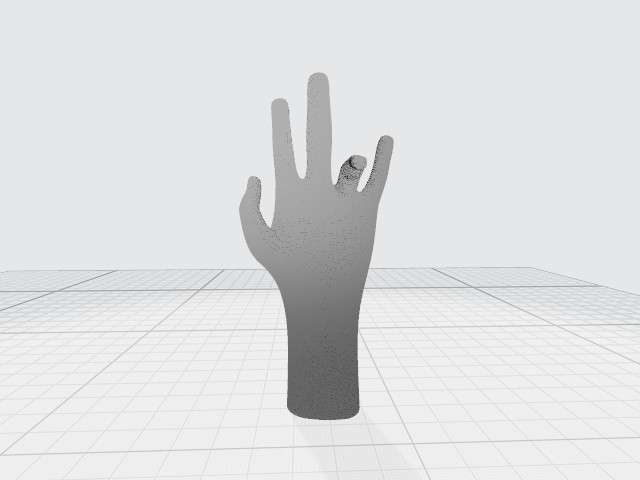}}
\label{figure:hand_7_90}

\caption{Sample 7 from the HandCT 3D dataset viewed from three different angles. In these views, one can see that the ring finger is bent.}
\label{figure:sample_7}
\end{figure*}

\begin{figure*}
\begin{adjustbox}{max width=\textwidth}
\begin{tabular}{M{5.5cm}M{5.5cm}M{5.5cm}}
Interpolation Method & Inferred Sinogram & Asolute difference with target  \\ 
\thickhline
Proposed &
\subfloat{\includegraphics[width=0.3\textwidth]{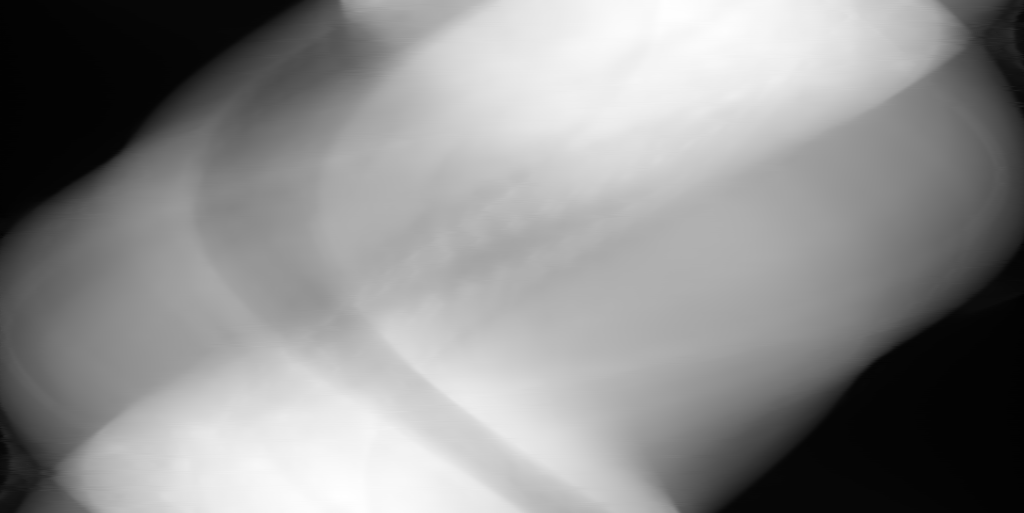}}\label{figure:inferred_sinogram_unet} &
\subfloat{\includegraphics[width=0.3\textwidth]{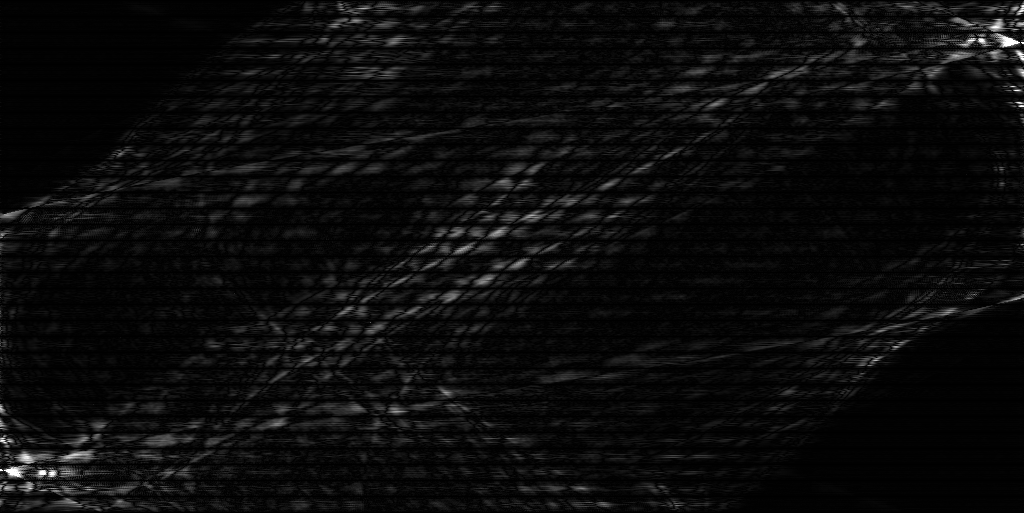}}\label{figure:diff_unet_sinogram_60} \\

Linear Interpolation &
\subfloat{\includegraphics[width=0.3\textwidth]{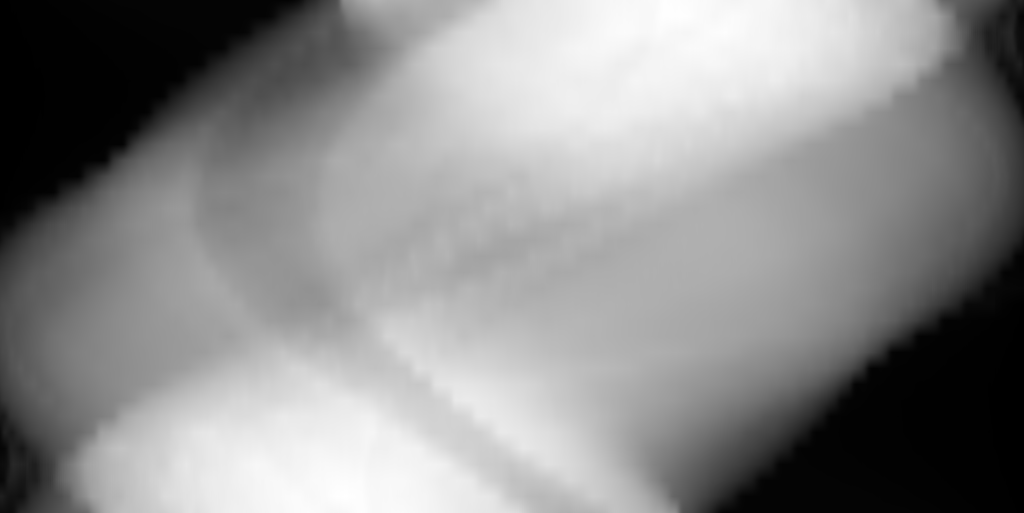}}\label{figure:inferred_sinogram_linear} &
\subfloat{\includegraphics[width=0.3\textwidth]{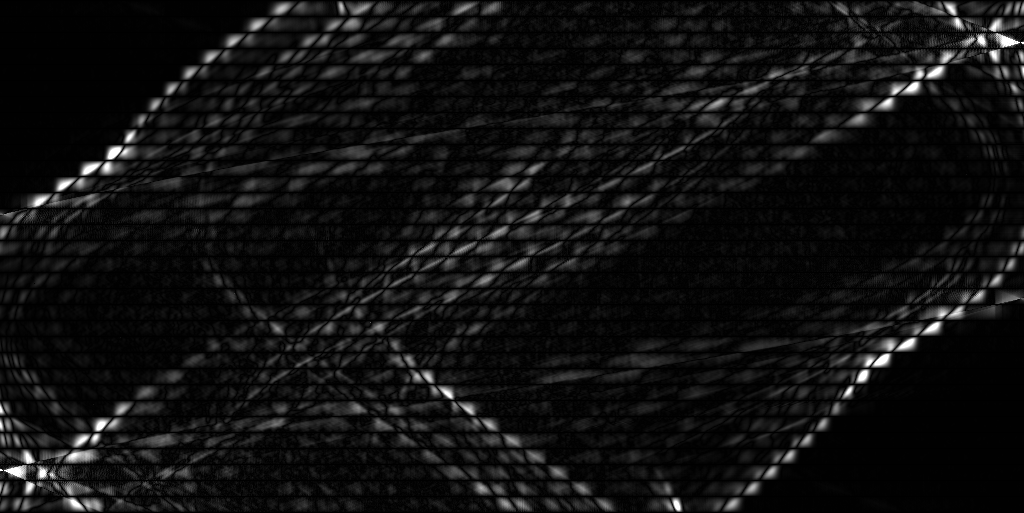}}\label{figure:diff_linear_sinogram_60}  \\
 
Bilinear Up-Sampling &
\subfloat{\includegraphics[width=0.3\textwidth]{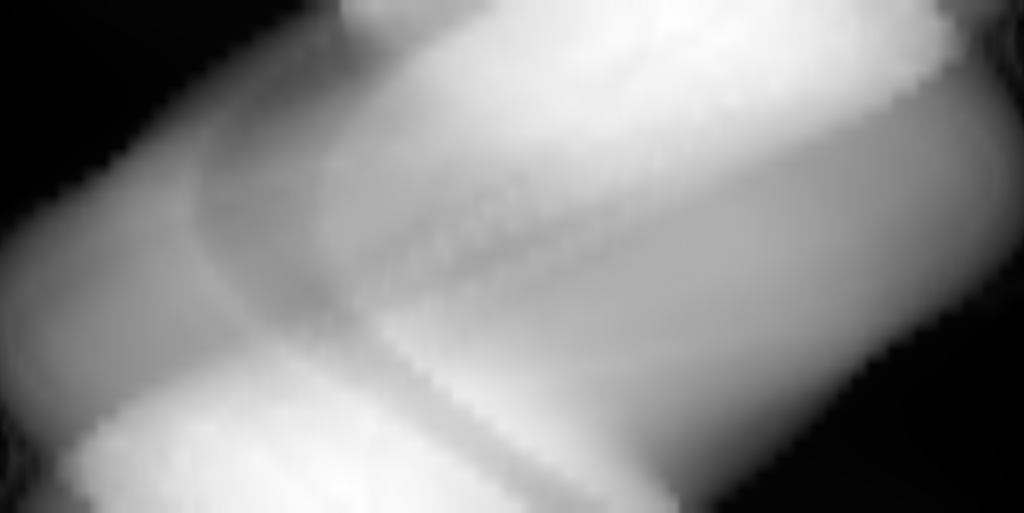}}\label{figure:inferred_sinogram_bilinear} &
\subfloat{\includegraphics[width=0.3\textwidth]{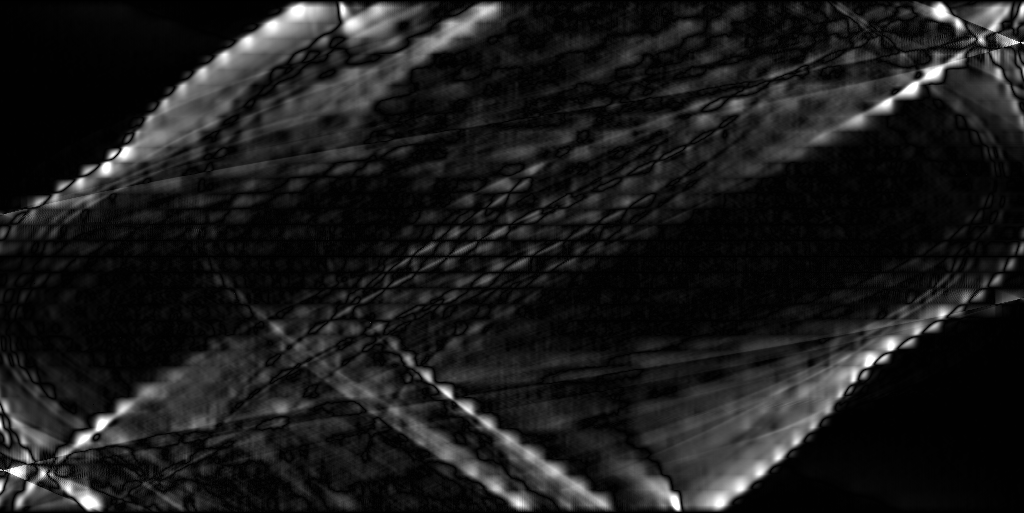}}\label{figure:diff_bilinear_sinogram_60}  \\

Nearest-Neighbours Up-Sampling &
\subfloat{\includegraphics[width=0.3\textwidth]{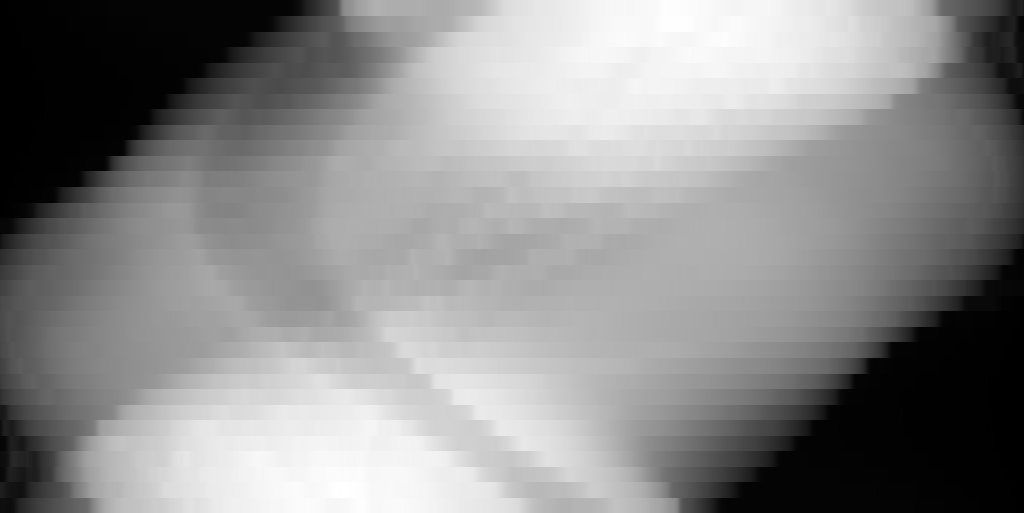}}\label{figure:inferred_sinogram_nearest} &
\subfloat{\includegraphics[width=0.3\textwidth]{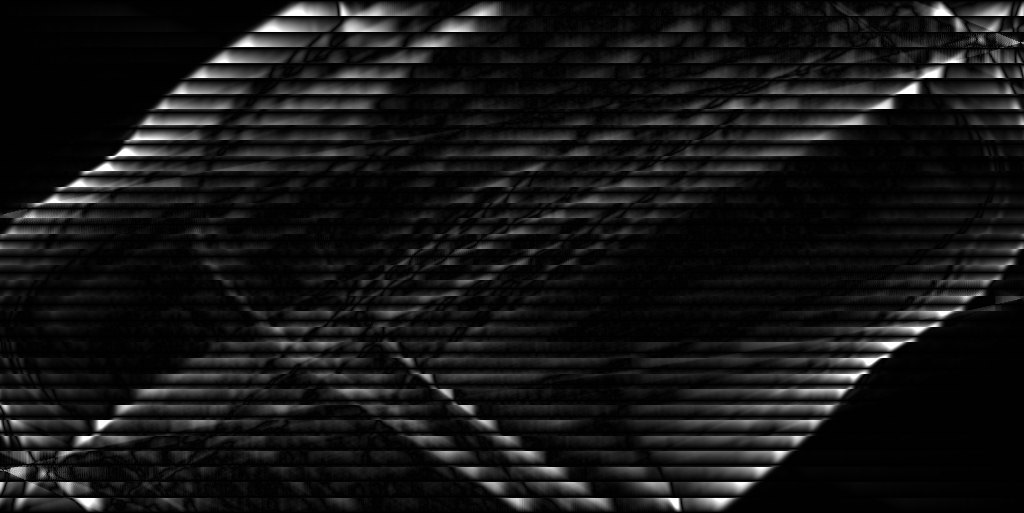}}\label{figure:diff_nearest_sinogram_60}  \\ 

\thickhline
\end{tabular}
\end{adjustbox}
\label{figure:sinograms_274_60}
\caption{Comparison of the inferred sinograms quality and their absolute differences with the target one for the \nth{60} slice of the patient 74. The main visible artefact corrected by our method is the aliasing of the sinogram. We think that a sequence of better acquisitions yields an image that looks smoother. For display purposes, we normalise the absolute difference between target and inferred sinograms, multiply it by 255 and then, to enhance contrast we used a gain of 2 and to enhance brighness we used a bias of 0.5.}
\end{figure*}

\begin{figure*}

\centering
\subfloat[Target]{\includegraphics[width=0.2\textwidth]{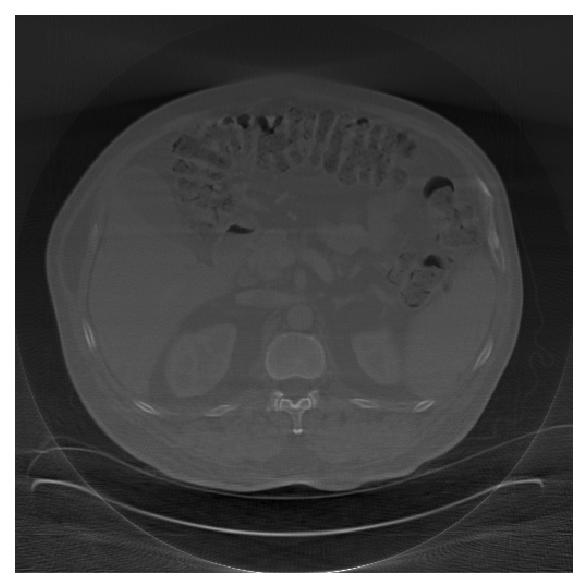}}
\label{figure:target}
\subfloat[Scarce]{\includegraphics[width=0.2\textwidth]{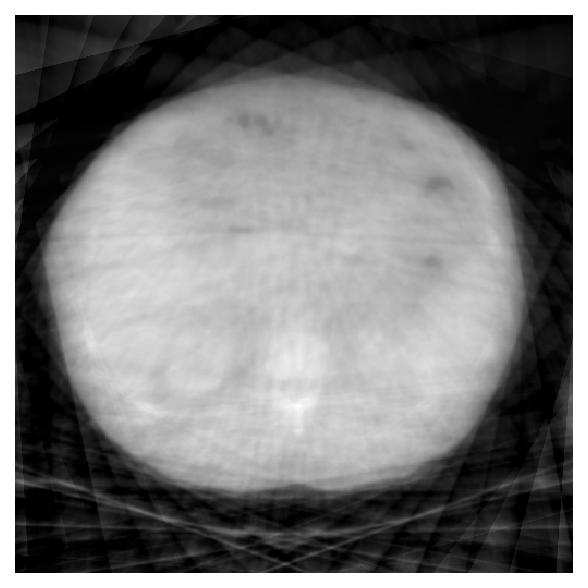}}
\label{figure:scarce}
\subfloat[Linear]{\includegraphics[width=0.2\textwidth]{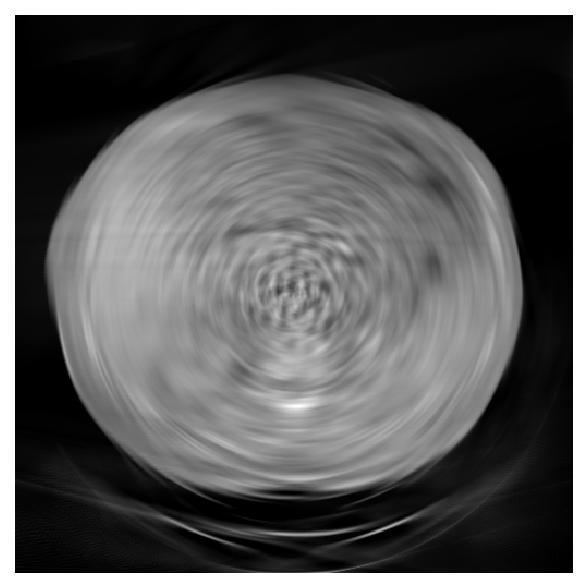}}
\label{figure:linearInterpolation}
\subfloat[Proposed]{\includegraphics[width=0.2\textwidth]{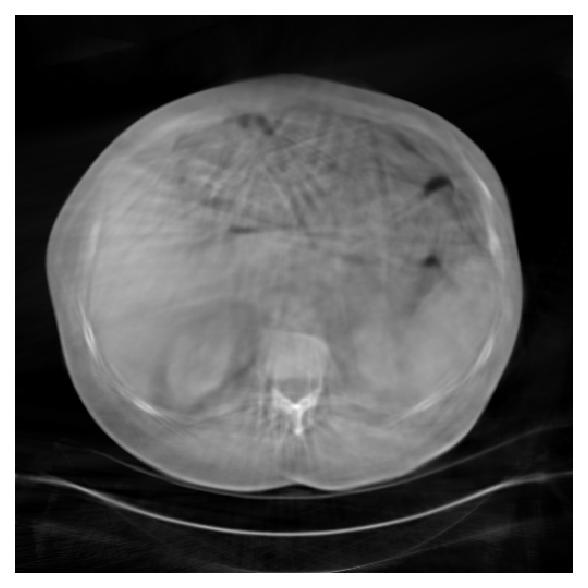}}
\label{figure:ours}

\caption{Reconstructions of the slice 60 of patient with Id 274, with the SART algorithm. The scarce sinogram is made of 33 acquisitions, and the target one of 33. It means that, given 33, 480 acquisitions are interpolated.}
\label{figure:LungCT_reconstructions}
\end{figure*}

\begin{figure*}
\begin{adjustbox}{max width=\textwidth}
\begin{tabular}{ccccc}
Interpolation Method & $\frac{1}{2}$ of acquisitions & $\frac{1}{4}$ of acquisitions & $\frac{1}{8}$ of acquisitions & $\frac{1}{16}$ of acquisitions \\ 
\thickhline
Proposed &
\subfloat{\includegraphics[width=0.3\textwidth]{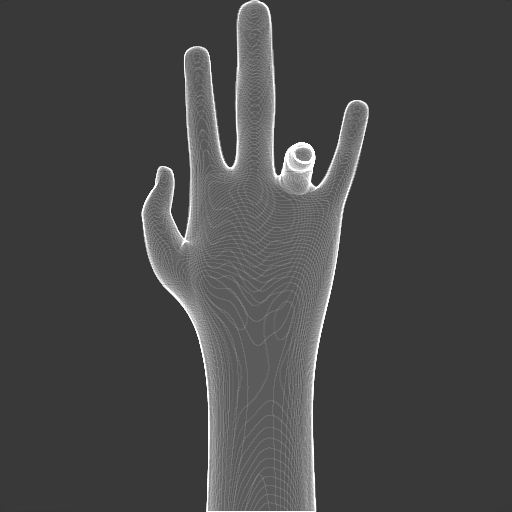}}\label{figure:unet_acquisition_263_2} &
\subfloat{\includegraphics[width=0.3\textwidth]{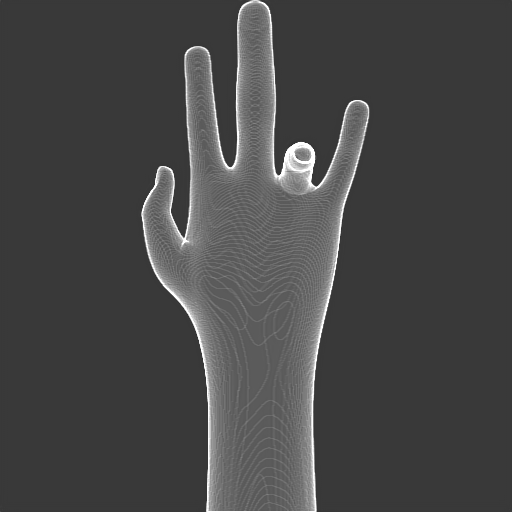}}\label{figure:unet_acquisition_262_4} &
\subfloat{\includegraphics[width=0.3\textwidth]{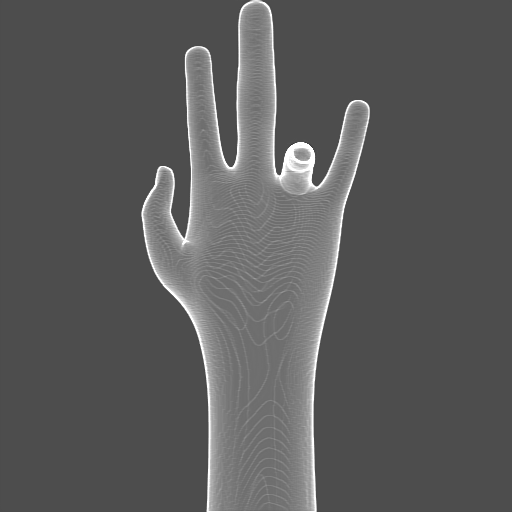}}\label{figure:unet_acquisition_260_8} &
\subfloat{\includegraphics[width=0.3\textwidth]{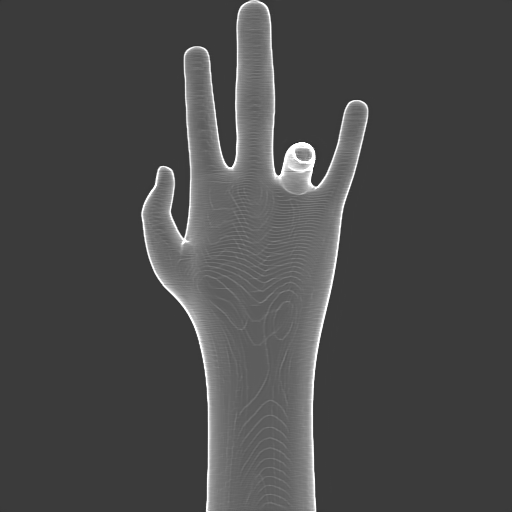}}\label{figure:unet_acquisition_264_16} \\

Linear Interpolation &
\subfloat{\includegraphics[width=0.3\textwidth]{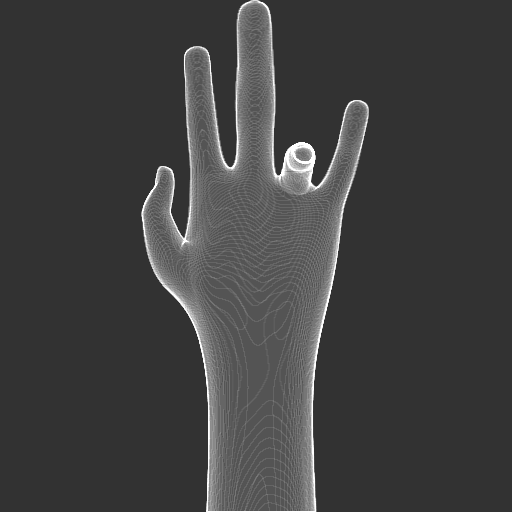}}\label{figure:linear_acquisition_263_2} &
\subfloat{\includegraphics[width=0.3\textwidth]{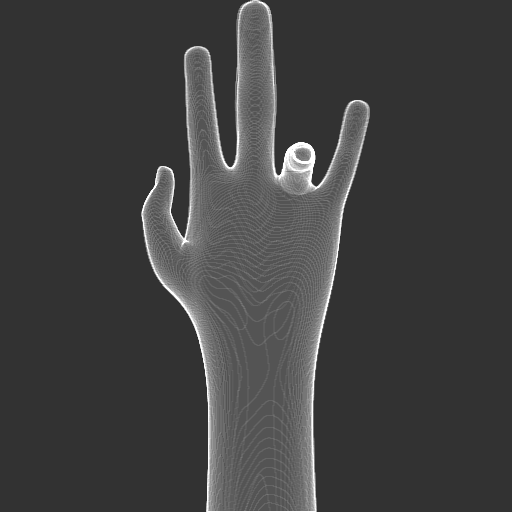}}\label{figure:linear_acquisition_262_4} &
\subfloat{\includegraphics[width=0.3\textwidth]{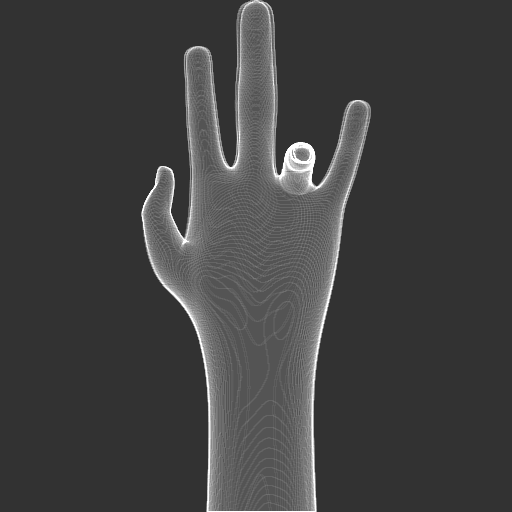}}\label{figure:linear_acquisition_260_8} &
\subfloat{\includegraphics[width=0.3\textwidth]{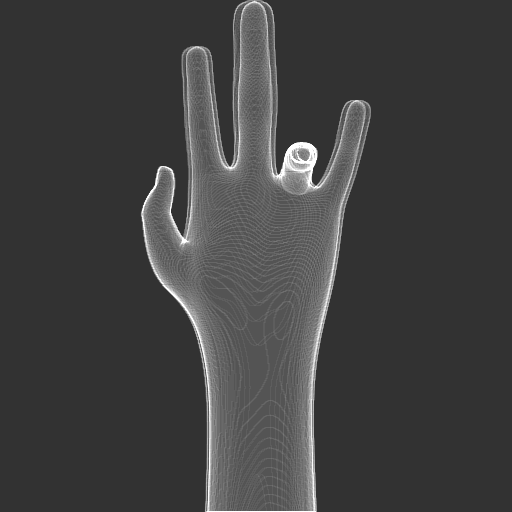}}\label{figure:linear_acquisition_264_16} \\
 
Trilinear Up-Sampling &
\subfloat{\includegraphics[width=0.3\textwidth]{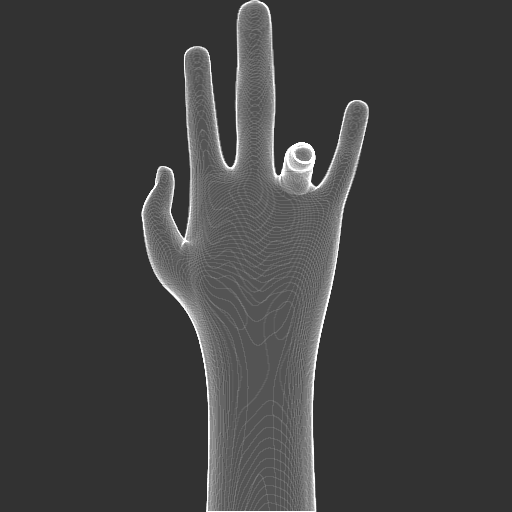}}\label{figure:trilinear_acquisition_263_2} &
\subfloat{\includegraphics[width=0.3\textwidth]{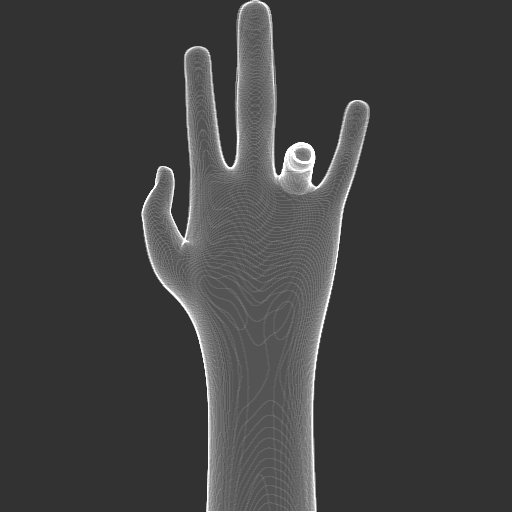}}\label{figure:trilinear_acquisition_262_4} &
\subfloat{\includegraphics[width=0.3\textwidth]{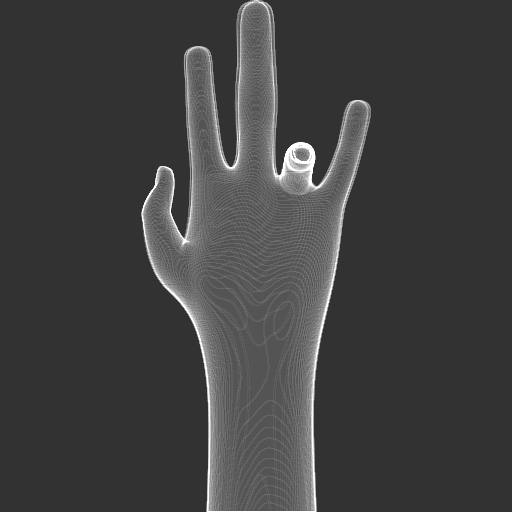}}\label{figure:trilinear_acquisition_260_8} &
\subfloat{\includegraphics[width=0.3\textwidth]{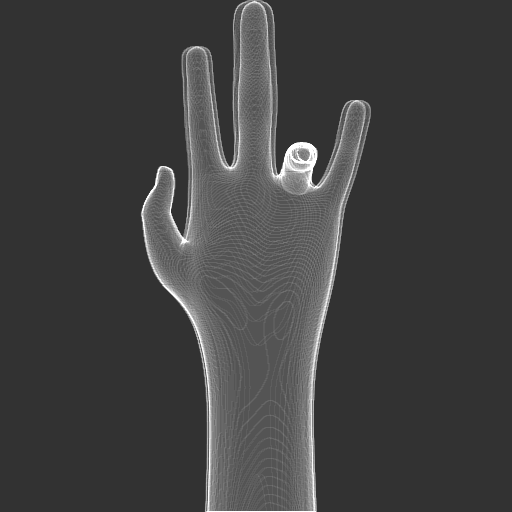}}\label{figure:trilinear_acquisition_264_16} \\

\thickhline
\end{tabular}
\end{adjustbox}
\label{figure:acquisitions_HandCT}
\caption{Comparison of the inferred acquisitions quality around angle 90\textdegree of the \nth{7} sample of the HandCT test dataset. For display purpose, the acquisitions are first normalised and multiplied by 255. Then, to enhance contrast we used a gain of 5 and to enhance brighness we used a bias of 50.}
\end{figure*}

%% file: sections/tables.tex
\begin{table*}[p]
\centering
\begin{tabular}{lccc}
\hline
Scanning Parameter          & Lung CT                & 3D Hands               \\
\hline
Beam geometry               & fan                    & parallel               \\

Detector size (pixels)      & 1024                   & 512*512                \\

Number of projections       & 513                    & 513                    \\

Object size (pixels)        & 512*512                & 512*512*512            \\
\hline
\end{tabular}
\label{table:scanning_geometries}
\caption{Scanning parameters used for each dataset.}
\end{table*}

\begin{table*}[p]
\centering
\begin{tabular}{lcccc}
\hline
&  Proposed &  Linear &  Bilinear &  Nearest-neighbours \\
\hline
146 &  \textbf{ 46.92 $\pm$ 0.86 } &  41.54 $\pm$ 0.55 &  37.95 $\pm$ 1.09 &  35.99 $\pm$ 1.17\\
274 &  \textbf{ 44.03 $\pm$ 1.86 } &  40.46 $\pm$ 1.77 &  35.62 $\pm$ 1.46 &  32.97 $\pm$ 1.39\\
127 &  \textbf{ 46.29 $\pm$ 0.87 } &  39.45 $\pm$ 0.96 &  35.79 $\pm$ 1.10 &  33.83 $\pm$ 1.16\\
64 &  \textbf{ 45.50 $\pm$ 0.61 } &  40.76 $\pm$ 0.44 &  36.98 $\pm$ 0.88 &  34.92 $\pm$ 1.07\\
143 &  \textbf{ 45.81 $\pm$ 0.91 } &  39.74 $\pm$ 0.56 &  36.00 $\pm$ 0.70 &  33.99 $\pm$ 0.83\\
\hline
\end{tabular}
\label{table:2D_LungCT_sinogram-PSNR}
\caption{Sinogram-PSNR of the LungCT test dataset, when enhanced using our proposed approach, linear interpolation, bilinear up-sampling and nearest-neighbours up-sampling. Up-sampling methods underperform compared to interpolation ones. The proposed approach yields on average a 5.4dB PSNR improvement compared to linear interpolation.}

\end{table*}

\begin{table*}[p]
\centering
\begin{tabular}{lccccc}
\hline
 &  Scarce &  Proposed &  Linear &  Bilinear&  Nearest-neighbours\\
\hline
146 &  39.45 $\pm$ 0.48 &  \textbf{ 63.61 $\pm$ 1.35 } &  57.96 $\pm$ 1.68 &  49.03 $\pm$ 3.41 &  48.79 $\pm$ 3.36\\
274 &  41.54 $\pm$ 0.75 &  \textbf{ 59.71 $\pm$ 2.53 } &  54.95 $\pm$ 2.45 &  45.07 $\pm$ 1.55 &  45.04 $\pm$ 1.49\\
127 &  40.13 $\pm$ 1.01 &  \textbf{ 62.99 $\pm$ 1.00 } &  55.36 $\pm$ 1.45 &  46.90 $\pm$ 2.79 &  46.61 $\pm$ 2.77\\
64 &  40.49 $\pm$ 0.63 &  \textbf{ 61.79 $\pm$ 0.93 } &  56.25 $\pm$ 1.70 &  47.96 $\pm$ 3.49 &  47.77 $\pm$ 3.45\\
143 &  40.26 $\pm$ 0.80 &  \textbf{ 62.42 $\pm$ 1.64 } &  55.39 $\pm$ 1.41 &  46.81 $\pm$ 2.67 &  46.72 $\pm$ 2.66\\
\hline
\end{tabular}
\label{table:2D_LungCT_reconstruction-PSNR}
\caption{Image-PSNR of the LungCT test dataset, when reconstructed from the scarce sinogram, and the ones enhanced using the methods of Table \ref{table:2D_LungCT_sinogram-PSNR}. Consistently with the results on the sinogram-PSNR, the roposed method improves by 6.1dB PSNR, on average, the quality of the reconstructed image.}
\end{table*}

\begin{table*}[p]
\centering
\begin{tabular}{lcccc}
\hline
&  Proposed &  Linear & Bilinear & Nearest-neighbours \\
\hline
146 &  \textbf{ 47.75 $\pm$ 0.87 } &  46.72 $\pm$ 0.72 &  45.43 $\pm$ 0.97 &  45.93 $\pm$ 0.63\\
274 &  \textbf{ 44.92 $\pm$ 1.84 } &  44.61 $\pm$ 1.76 &  42.30 $\pm$ 1.70 &  43.77 $\pm$ 1.43\\
127 &  \textbf{ 47.17 $\pm$ 0.91 } &  45.30 $\pm$ 0.93 &  43.90 $\pm$ 1.13 &  44.67 $\pm$ 0.80\\
64 &  \textbf{ 46.29 $\pm$ 0.63 } &  45.08 $\pm$ 0.69 &  43.72 $\pm$ 0.58 &  44.37 $\pm$ 0.71\\
143 &  \textbf{ 46.69 $\pm$ 0.96 } &  44.65 $\pm$ 0.64 &  43.32 $\pm$ 0.71 &  44.06 $\pm$ 0.51\\
\hline
\end{tabular}
\label{table:2D_LungCT_sinogram-PSNR_improved}
\caption{Sinogram-PSNR for the test samples of the LungCT dataset, when enhanced using the methods of Table \ref{table:2D_LungCT_sinogram-PSNR}, and improved with an image processing technique. The performance gain is noticeable for deterministic methods, but less significant for the proposed data-driven approach.}
\end{table*}

\begin{table*}[p]
\centering
\begin{tabular}{lcccc}
\hline
 &  Proposed &  Linear &  Bilinear  &  Nearest-neighbours \\
\hline
146  &  \textbf{ 64.09 $\pm$ 1.37 } &  60.83 $\pm$ 0.74 &  56.17 $\pm$ 3.35 &  54.04 $\pm$ 0.54\\
274  &  \textbf{ 60.72 $\pm$ 2.55 } &  58.58 $\pm$ 1.80 &  52.28 $\pm$ 2.92 &  54.57 $\pm$ 0.68\\
127  &  \textbf{ 63.78 $\pm$ 1.05 } &  59.54 $\pm$ 0.84 &  54.28 $\pm$ 3.22 &  54.59 $\pm$ 0.93\\
64  &  \textbf{ 62.41 $\pm$ 0.94 } &  59.30 $\pm$ 0.71 &  54.42 $\pm$ 2.83 &  54.00 $\pm$ 0.86\\
143  &  \textbf{ 63.13 $\pm$ 1.71 } &  58.91 $\pm$ 0.96 &  54.10 $\pm$ 2.76 &  54.02 $\pm$ 0.69\\
\hline
\end{tabular}
\label{table:2D_LungCT_reconstruction-PSNR_improved}
\caption{Image-PSNR of the LungCT test dataset, when reconstructed from the scarce sinogram, and the ones enhanced using the methods of Table \ref{table:2D_LungCT_sinogram-PSNR_improved}.}
\end{table*}

\begin{table*}[]
\centering
\begin{tabular}{lccc}
\hline
 &  Proposed &  Linear Interpolation &  Trilinear Up-sampling\\
\hline
7 &  \textbf{ 49.05} &  42.72 &  41.40\\
18 &  \textbf{ 49.11} &  42.22 &  40.87\\
3 &  \textbf{ 48.04} &  41.58 &  40.42\\
2 &  \textbf{ 49.05} &  42.73 &  41.22\\
46 &  \textbf{ 49.39} &  42.98 &  41.42\\
\hline
\end{tabular}
\label{table:3D_handCT_sinogram-PSNR}
\caption{Sinogram-PSNR of the HandCT test dataset, when enhanced using our proposed approach, linear interpolation and trilinear up-sampling. On average, 7.86dB PSNR are gained when using our approach, compared to a linear interpolation of the measurements.}
\end{table*}

\begin{table*}[]
\centering
\begin{tabular}{lcccc}
\hline
 &  Scarce &  Proposed &  Linear Interpolation &  Trilinear Up-sampling\\
\hline
7 &  51.64 &  \textbf{ 73.16} &  65.94 &  62.94\\
18 &  51.62 &  \textbf{ 73.01} &  64.83 &  61.67\\
3 &  51.68 &  \textbf{ 70.66} &  63.73 &  61.02\\
2 &  51.70 &  \textbf{ 72.57} &  65.48 &  62.04\\
46 &  51.64 &  \textbf{ 73.47} &  66.00 &  62.39\\
\hline
\end{tabular}
\label{table:3D_handCT_reconstruction-PSNR}
\caption{Image-PSNR of the HandCT test dataset, when reconstructed from the scarce sinogram, and the ones enhanced using the methods of Table 6. The gain of PSNR is here of 7.4dB, on average.}
\end{table*}

\begin{table*}

\centering
\begin{tabular}{cccccccccccccccccc}
Acquisition indexes  & 0 & 1 & 2 & 3 & 4 & 5 & 6 & 7 & 8 & 9 & 10 & 11 & 12 & 13 & 14 & 15 & 16 \\
\hline
$\frac{1}{16}$ of acquisitions & 0 &\sout{1} & \sout{2} & \sout{3} & \sout{4} & \sout{5} & \sout{6} & \sout{7} & \circled{8} & \sout{9} & \sout{10} & \sout{11} & \sout{12} & \sout{13} & \sout{14} & \sout{15} & 16 \\
$\frac{1}{8}$ of acquisitions & 0 &\sout{1} & \sout{2} & \sout{3} & \circled{4} & \sout{5} & \sout{6} & \sout{7} & 8 & \sout{9} & \sout{10} & \sout{11} & \sout{12} & \sout{13} & \sout{14} & \sout{15} & 16 \\
$\frac{1}{4}$ of acquisitions & 0 &\sout{1} & \sout{2} & \sout{3} & 4 & \sout{5} & \circled{6} & \sout{7} & 8 & \sout{9} & \sout{10} & \sout{11} & 12 & \sout{13} & \sout{14} & \sout{15} & 16 \\
$\frac{1}{2}$ of acquisitions & 0 & \sout{1} & 2 & \sout{3} & 4 & \sout{5} & 6 & \circled{7} & 8 & \sout{9} & 10 & \sout{11} & 12 & \sout{13} & 14 & \sout{15} & 16 \\
\end{tabular}
\caption{Acquisitions indexes available and missing (struck-through) for a given proportion of acquisitions. In \ref{figure:acquisitions_HandCT}, we chose to plot a different inferred (circled)  acquisition for each up-sampling ratio. This choice was made to keep an angular proximity between them (they are all in an interval of 3.5\textdegree) and so that they would all be at mid-distance between reference acquisitions.}
\label{table:angles_available_and_missing}
\end{table*}

%% file: sections/figures.tex
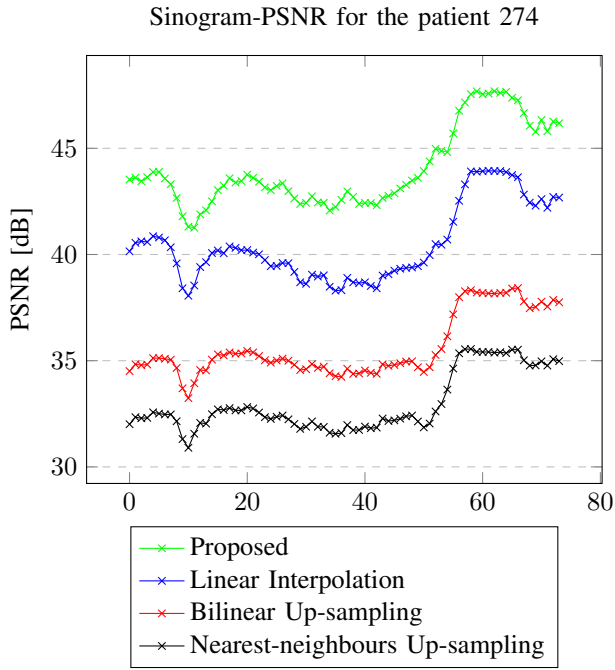
\begin{figure}
\centering
\begin{tikzpicture}
\begin{axis}
[title={Sinogram-PSNR for the patient 274},xlabel={Slice Index},ylabel={PSNR [dB]},legend style={at={(0.5,-0.1)},anchor=north},ymajorgrids=true,grid style=dashed,legend cell align={left},]
    
\addplot[color=green, mark=x]
coordinates { (0,43.53)(1,43.62)(2,43.45)(3,43.64)(4,43.88)(5,43.89)(6,43.58)(7,43.31)(8,42.66)(9,41.80)(10,41.31)(11,41.26)(12,41.89)(13,42.08)(14,42.50)(15,43.03)(16,43.24)(17,43.59)(18,43.39)(19,43.46)(20,43.76)(21,43.61)(22,43.43)(23,43.16)(24,43.03)(25,43.23)(26,43.35)(27,42.98)(28,42.64)(29,42.38)(30,42.43)(31,42.74)(32,42.45)(33,42.44)(34,42.07)(35,42.24)(36,42.57)(37,42.97)(38,42.71)(39,42.38)(40,42.43)(41,42.42)(42,42.32)(43,42.66)(44,42.76)(45,42.88)(46,43.11)(47,43.28)(48,43.48)(49,43.61)(50,43.93)(51,44.38)(52,44.98)(53,44.88)(54,44.82)(55,45.70)(56,46.77)(57,47.16)(58,47.55)(59,47.68)(60,47.55)(61,47.58)(62,47.69)(63,47.61)(64,47.66)(65,47.38)(66,47.26)(67,46.66)(68,46.06)(69,45.78)(70,46.33)(71,45.81)(72,46.26)(73,46.18) };
    
\addplot[color=blue, mark=x]
coordinates { (0,40.15)(1,40.55)(2,40.62)(3,40.59)(4,40.86)(5,40.80)(6,40.67)(7,40.34)(8,39.58)(9,38.41)(10,38.06)(11,38.54)(12,39.41)(13,39.64)(14,40.07)(15,40.19)(16,40.06)(17,40.38)(18,40.31)(19,40.20)(20,40.21)(21,40.09)(22,40.01)(23,39.74)(24,39.45)(25,39.46)(26,39.60)(27,39.59)(28,39.18)(29,38.69)(30,38.62)(31,39.05)(32,38.96)(33,39.02)(34,38.48)(35,38.30)(36,38.32)(37,38.90)(38,38.69)(39,38.65)(40,38.69)(41,38.52)(42,38.41)(43,39.01)(44,39.07)(45,39.24)(46,39.33)(47,39.37)(48,39.39)(49,39.45)(50,39.62)(51,39.97)(52,40.49)(53,40.46)(54,40.70)(55,41.54)(56,42.53)(57,43.30)(58,43.91)(59,43.90)(60,43.92)(61,43.94)(62,43.93)(63,43.93)(64,43.88)(65,43.74)(66,43.64)(67,42.82)(68,42.44)(69,42.30)(70,42.61)(71,42.20)(72,42.71)(73,42.68) };
    
\addplot[color=red, mark=x]
coordinates { (0,34.51)(1,34.83)(2,34.79)(3,34.82)(4,35.12)(5,35.12)(6,35.09)(7,35.05)(8,34.66)(9,33.70)(10,33.24)(11,33.94)(12,34.56)(13,34.54)(14,35.05)(15,35.30)(16,35.25)(17,35.39)(18,35.33)(19,35.34)(20,35.46)(21,35.39)(22,35.22)(23,35.02)(24,34.91)(25,35.01)(26,35.10)(27,35.00)(28,34.81)(29,34.56)(30,34.60)(31,34.84)(32,34.66)(33,34.72)(34,34.41)(35,34.27)(36,34.24)(37,34.62)(38,34.38)(39,34.40)(40,34.54)(41,34.43)(42,34.38)(43,34.83)(44,34.76)(45,34.80)(46,34.88)(47,34.96)(48,34.97)(49,34.69)(50,34.47)(51,34.69)(52,35.27)(53,35.55)(54,36.15)(55,37.18)(56,37.99)(57,38.27)(58,38.32)(59,38.21)(60,38.19)(61,38.17)(62,38.16)(63,38.19)(64,38.20)(65,38.40)(66,38.42)(67,37.79)(68,37.47)(69,37.53)(70,37.78)(71,37.55)(72,37.86)(73,37.75) };
    
\addplot[color=black, mark=x]
coordinates { (0,32.01)(1,32.34)(2,32.29)(3,32.31)(4,32.57)(5,32.51)(6,32.47)(7,32.47)(8,32.15)(9,31.31)(10,30.90)(11,31.56)(12,32.07)(13,32.04)(14,32.48)(15,32.71)(16,32.68)(17,32.77)(18,32.66)(19,32.66)(20,32.82)(21,32.75)(22,32.56)(23,32.34)(24,32.26)(25,32.36)(26,32.42)(27,32.24)(28,32.02)(29,31.79)(30,31.89)(31,32.14)(32,31.89)(33,31.88)(34,31.61)(35,31.56)(36,31.59)(37,31.97)(38,31.73)(39,31.74)(40,31.90)(41,31.82)(42,31.85)(43,32.28)(44,32.16)(45,32.19)(46,32.27)(47,32.38)(48,32.42)(49,32.14)(50,31.86)(51,32.04)(52,32.61)(53,32.96)(54,33.64)(55,34.63)(56,35.35)(57,35.54)(58,35.55)(59,35.42)(60,35.41)(61,35.41)(62,35.38)(63,35.40)(64,35.36)(65,35.52)(66,35.52)(67,35.02)(68,34.77)(69,34.78)(70,34.97)(71,34.77)(72,35.07)(73,34.98) };
\legend{Proposed,Linear Interpolation,Bilinear Up-sampling,Nearest-neighbours Up-sampling}
\end{axis}
\end{tikzpicture}
\caption{Per-slice sinogram-PSNR of the sample 274 of the LungCT test dataset. Interestingly, the deviation between interpolation quality is constant accross slice.}
\label{figure:sinogram_PSNR_274}
\end{figure}

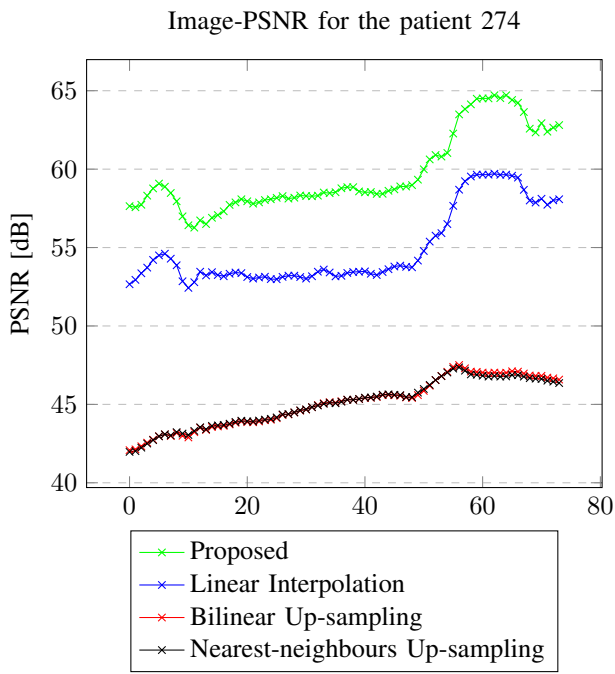
\begin{figure}
\centering
\begin{tikzpicture}
\begin{axis}
[title={Image-PSNR for the patient 274},xlabel={Slice Index},ylabel={PSNR [dB]},legend style={at={(0.5,-0.1)},anchor=north},ymajorgrids=true,grid style=dashed,legend cell align={left},]
    
\addplot[color=green, mark=x]
coordinates { (0,57.64)(1,57.57)(2,57.73)(3,58.31)(4,58.77)(5,59.08)(6,58.88)(7,58.48)(8,57.95)(9,56.99)(10,56.42)(11,56.27)(12,56.73)(13,56.50)(14,56.90)(15,57.07)(16,57.31)(17,57.72)(18,57.88)(19,58.08)(20,57.95)(21,57.80)(22,57.88)(23,58.03)(24,58.08)(25,58.17)(26,58.28)(27,58.12)(28,58.18)(29,58.32)(30,58.29)(31,58.27)(32,58.31)(33,58.51)(34,58.47)(35,58.53)(36,58.78)(37,58.85)(38,58.85)(39,58.58)(40,58.52)(41,58.54)(42,58.42)(43,58.43)(44,58.62)(45,58.71)(46,58.90)(47,58.89)(48,58.98)(49,59.33)(50,59.99)(51,60.62)(52,60.90)(53,60.80)(54,61.03)(55,62.27)(56,63.49)(57,63.83)(58,64.13)(59,64.49)(60,64.51)(61,64.51)(62,64.71)(63,64.53)(64,64.71)(65,64.41)(66,64.23)(67,63.64)(68,62.58)(69,62.34)(70,62.93)(71,62.39)(72,62.65)(73,62.81) };
    
\addplot[color=blue, mark=x]
coordinates { (0,52.66)(1,52.94)(2,53.36)(3,53.72)(4,54.21)(5,54.50)(6,54.61)(7,54.29)(8,53.87)(9,52.85)(10,52.43)(11,52.79)(12,53.46)(13,53.22)(14,53.43)(15,53.25)(16,53.17)(17,53.33)(18,53.41)(19,53.37)(20,53.12)(21,53.02)(22,53.09)(23,53.14)(24,52.99)(25,52.98)(26,53.13)(27,53.21)(28,53.21)(29,53.12)(30,53.02)(31,53.17)(32,53.46)(33,53.60)(34,53.41)(35,53.16)(36,53.20)(37,53.38)(38,53.44)(39,53.46)(40,53.48)(41,53.33)(42,53.25)(43,53.44)(44,53.62)(45,53.79)(46,53.85)(47,53.77)(48,53.74)(49,54.17)(50,54.78)(51,55.38)(52,55.76)(53,55.92)(54,56.50)(55,57.66)(56,58.69)(57,59.23)(58,59.53)(59,59.65)(60,59.65)(61,59.65)(62,59.71)(63,59.64)(64,59.65)(65,59.58)(66,59.46)(67,58.68)(68,58.00)(69,57.87)(70,58.11)(71,57.73)(72,58.01)(73,58.08) };
    
\addplot[color=red, mark=x]
coordinates { (0,42.08)(1,42.13)(2,42.33)(3,42.56)(4,42.75)(5,42.97)(6,43.09)(7,42.98)(8,43.17)(9,43.00)(10,42.89)(11,43.23)(12,43.51)(13,43.37)(14,43.56)(15,43.59)(16,43.63)(17,43.72)(18,43.83)(19,43.91)(20,43.87)(21,43.86)(22,43.91)(23,43.99)(24,44.02)(25,44.12)(26,44.33)(27,44.36)(28,44.48)(29,44.60)(30,44.67)(31,44.82)(32,44.96)(33,45.07)(34,45.14)(35,45.11)(36,45.20)(37,45.30)(38,45.29)(39,45.34)(40,45.41)(41,45.42)(42,45.47)(43,45.59)(44,45.60)(45,45.58)(46,45.55)(47,45.44)(48,45.41)(49,45.59)(50,45.86)(51,46.19)(52,46.55)(53,46.81)(54,47.08)(55,47.39)(56,47.51)(57,47.30)(58,47.09)(59,47.07)(60,47.03)(61,46.99)(62,47.03)(63,47.00)(64,47.02)(65,47.11)(66,47.09)(67,46.97)(68,46.84)(69,46.82)(70,46.82)(71,46.71)(72,46.66)(73,46.58) };
    
\addplot[color=black, mark=x]
coordinates { (0,41.97)(1,42.02)(2,42.24)(3,42.50)(4,42.72)(5,42.96)(6,43.09)(7,43.02)(8,43.23)(9,43.14)(10,43.05)(11,43.31)(12,43.54)(13,43.41)(14,43.63)(15,43.66)(16,43.69)(17,43.77)(18,43.88)(19,43.96)(20,43.92)(21,43.91)(22,43.97)(23,44.03)(24,44.08)(25,44.17)(26,44.36)(27,44.38)(28,44.49)(29,44.62)(30,44.68)(31,44.82)(32,44.93)(33,45.04)(34,45.11)(35,45.07)(36,45.17)(37,45.30)(38,45.29)(39,45.35)(40,45.43)(41,45.44)(42,45.50)(43,45.62)(44,45.64)(45,45.61)(46,45.59)(47,45.47)(48,45.46)(49,45.75)(50,45.98)(51,46.24)(52,46.55)(53,46.80)(54,47.03)(55,47.28)(56,47.36)(57,47.15)(58,46.91)(59,46.89)(60,46.83)(61,46.77)(62,46.81)(63,46.77)(64,46.78)(65,46.87)(66,46.86)(67,46.77)(68,46.67)(69,46.64)(70,46.62)(71,46.51)(72,46.45)(73,46.36) };
\legend{Proposed,Linear Interpolation,Bilinear Up-sampling,Nearest-neighbours Up-sampling}
\end{axis}

\end{tikzpicture}
\caption{Per-slice image-PSNR of the sample 274 of the LungCT test dataset. Even though in the sinogram space, inferred acquisitions quality differs accross up-sampling methods, the difference is less significant in the image space.}
\label{figure:image_PSNR_274}
\end{figure}

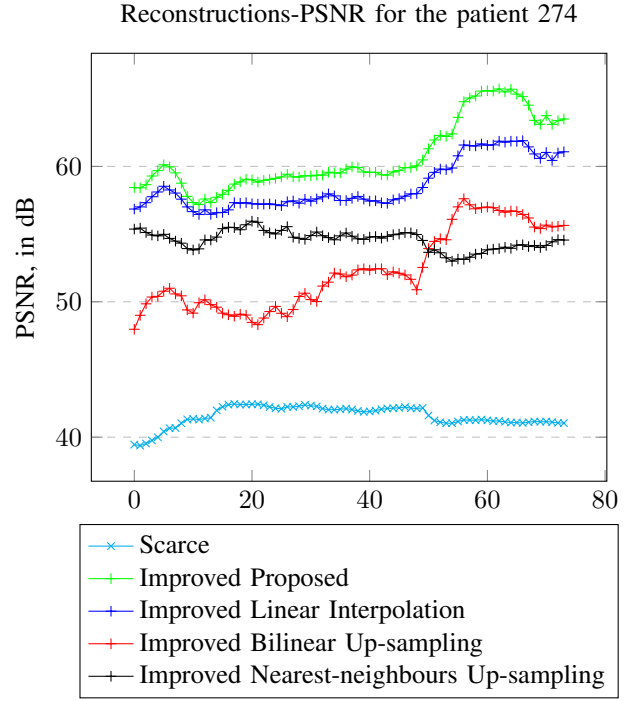
\begin{figure}
\centering
\begin{tikzpicture}
\begin{axis}
[title={Reconstructions-PSNR for the patient 274},xlabel={Slice Index},ylabel={PSNR, in dB},legend style={at={(0.5,-0.1)},anchor=north},ymajorgrids=true,grid style=dashed,legend cell align={left},]
    
\addplot[color=cyan, mark=x]
coordinates { (0,39.45)(1,39.38)(2,39.53)(3,39.76)(4,39.99)(5,40.42)(6,40.66)(7,40.68)(8,41.03)(9,41.31)(10,41.35)(11,41.30)(12,41.38)(13,41.44)(14,41.98)(15,42.24)(16,42.40)(17,42.45)(18,42.41)(19,42.41)(20,42.43)(21,42.43)(22,42.36)(23,42.23)(24,42.13)(25,42.10)(26,42.24)(27,42.23)(28,42.29)(29,42.39)(30,42.33)(31,42.27)(32,42.11)(33,42.03)(34,42.03)(35,42.06)(36,42.11)(37,42.04)(38,41.94)(39,41.87)(40,41.90)(41,41.96)(42,42.06)(43,42.12)(44,42.14)(45,42.17)(46,42.22)(47,42.14)(48,42.12)(49,42.16)(50,41.60)(51,41.23)(52,41.10)(53,41.03)(54,41.03)(55,41.16)(56,41.27)(57,41.26)(58,41.25)(59,41.30)(60,41.24)(61,41.19)(62,41.20)(63,41.13)(64,41.08)(65,41.08)(66,41.05)(67,41.10)(68,41.16)(69,41.14)(70,41.15)(71,41.11)(72,41.06)(73,41.04) };
    
\addplot[color=green, mark=+]
coordinates { (0,58.44)(1,58.42)(2,58.65)(3,59.29)(4,59.69)(5,60.12)(6,59.96)(7,59.52)(8,58.76)(9,57.79)(10,57.33)(11,57.18)(12,57.59)(13,57.33)(14,57.72)(15,57.94)(16,58.23)(17,58.69)(18,58.86)(19,59.05)(20,59.01)(21,58.84)(22,58.97)(23,59.04)(24,59.12)(25,59.22)(26,59.42)(27,59.23)(28,59.21)(29,59.32)(30,59.30)(31,59.35)(32,59.35)(33,59.57)(34,59.52)(35,59.52)(36,59.80)(37,59.96)(38,59.91)(39,59.59)(40,59.56)(41,59.56)(42,59.38)(43,59.36)(44,59.59)(45,59.69)(46,59.91)(47,59.91)(48,60.04)(49,60.48)(50,61.30)(51,61.95)(52,62.29)(53,62.21)(54,62.40)(55,63.62)(56,64.78)(57,65.05)(58,65.20)(59,65.56)(60,65.59)(61,65.56)(62,65.72)(63,65.50)(64,65.70)(65,65.34)(66,65.15)(67,64.52)(68,63.40)(69,63.11)(70,63.75)(71,63.10)(72,63.38)(73,63.50) };
    
\addplot[color=blue, mark=+]
coordinates { (0,56.85)(1,57.01)(2,57.32)(3,57.76)(4,58.11)(5,58.53)(6,58.26)(7,58.03)(8,57.58)(9,57.01)(10,56.66)(11,56.45)(12,56.79)(13,56.46)(14,56.58)(15,56.60)(16,56.79)(17,57.32)(18,57.28)(19,57.32)(20,57.24)(21,57.22)(22,57.21)(23,57.23)(24,57.17)(25,57.10)(26,57.39)(27,57.45)(28,57.28)(29,57.57)(30,57.43)(31,57.60)(32,57.75)(33,57.97)(34,57.82)(35,57.48)(36,57.48)(37,57.64)(38,57.78)(39,57.58)(40,57.44)(41,57.46)(42,57.31)(43,57.27)(44,57.53)(45,57.62)(46,57.80)(47,57.97)(48,57.99)(49,58.44)(50,59.13)(51,59.56)(52,59.80)(53,59.76)(54,59.88)(55,60.78)(56,61.58)(57,61.53)(58,61.51)(59,61.66)(60,61.56)(61,61.59)(62,61.86)(63,61.79)(64,61.86)(65,61.84)(66,61.90)(67,61.44)(68,60.92)(69,60.59)(70,61.03)(71,60.44)(72,60.99)(73,61.08) };
    
\addplot[color=red, mark=+]
coordinates { (0,47.97)(1,49.00)(2,49.85)(3,50.35)(4,50.40)(5,50.78)(6,51.00)(7,50.58)(8,50.44)(9,49.40)(10,49.16)(11,49.93)(12,50.15)(13,49.79)(14,49.60)(15,49.16)(16,49.04)(17,48.93)(18,49.08)(19,49.02)(20,48.48)(21,48.31)(22,48.80)(23,49.28)(24,49.66)(25,49.15)(26,48.90)(27,49.43)(28,50.39)(29,50.63)(30,50.16)(31,50.02)(32,51.16)(33,51.45)(34,52.12)(35,52.05)(36,51.85)(37,51.97)(38,52.36)(39,52.44)(40,52.34)(41,52.44)(42,52.44)(43,52.00)(44,52.21)(45,52.11)(46,51.98)(47,51.66)(48,50.89)(49,52.54)(50,53.93)(51,54.46)(52,54.66)(53,54.58)(54,56.09)(55,57.01)(56,57.61)(57,57.16)(58,56.87)(59,56.95)(60,57.00)(61,56.95)(62,56.74)(63,56.62)(64,56.71)(65,56.67)(66,56.45)(67,56.19)(68,55.52)(69,55.41)(70,55.66)(71,55.53)(72,55.58)(73,55.64) };
    
\addplot[color=black, mark=+]
coordinates { (0,55.37)(1,55.44)(2,55.12)(3,54.93)(4,54.87)(5,54.98)(6,54.68)(7,54.48)(8,54.33)(9,53.93)(10,53.83)(11,53.91)(12,54.57)(13,54.56)(14,54.78)(15,55.40)(16,55.50)(17,55.48)(18,55.32)(19,55.69)(20,55.95)(21,55.87)(22,55.27)(23,55.10)(24,55.01)(25,55.27)(26,55.55)(27,54.77)(28,54.69)(29,54.61)(30,54.89)(31,55.15)(32,54.88)(33,54.74)(34,54.58)(35,54.87)(36,55.08)(37,54.84)(38,54.65)(39,54.61)(40,54.78)(41,54.82)(42,54.70)(43,54.83)(44,54.95)(45,55.00)(46,55.11)(47,55.09)(48,54.97)(49,54.53)(50,53.65)(51,53.84)(52,53.62)(53,53.25)(54,53.00)(55,53.16)(56,53.13)(57,53.29)(58,53.52)(59,53.57)(60,53.84)(61,53.89)(62,53.93)(63,54.03)(64,53.95)(65,54.18)(66,54.21)(67,54.08)(68,54.16)(69,54.01)(70,54.16)(71,54.43)(72,54.59)(73,54.56) };
\legend{Scarce,Improved Proposed,Improved Linear Interpolation,Improved Bilinear Up-sampling,Improved Nearest-neighbours Up-sampling}
\end{axis}

\end{tikzpicture}
\caption{Per-slice image-PSNR of the sample 274 of the LungCT test dataset, after 2D enhancement. The PSNR improvement is significant when using interpolation rather than up-sampling.}
\label{figure:image_PSNR_LungCT_improved}
\end{figure}

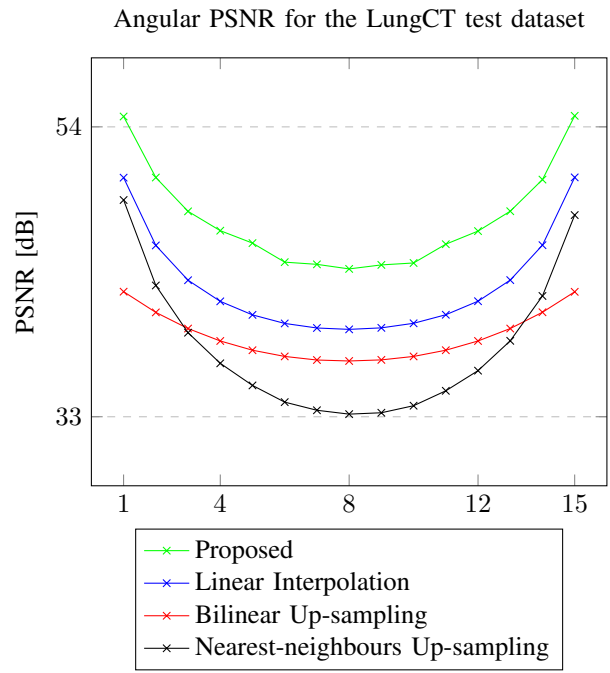
\begin{figure}
\centering
\begin{tikzpicture}
\begin{axis}[
title={Angular PSNR for the LungCT test dataset},
xlabel={Acquisition Index},
ylabel={PSNR [dB]},
xmin=0, xmax=16,
ymin=28, ymax=59,
xtick={1,4,8,12,15},
ytick={33, 54},
legend style={at={(0.5,-0.1)},anchor=north},
ymajorgrids=true,
grid style=dashed,
legend cell align={left},
]
    
\addplot[
color=green,
mark=x
]
coordinates { (1,54.75)(2,50.33)(3,47.88)(4,46.48)(5,45.58)(6,44.20)(7,44.04)(8,43.71)(9,44.00)(10,44.14)(11,45.50)(12,46.45)(13,47.89)(14,50.18)(15,54.80) };
    
\addplot[
color=blue,
mark=x
]
coordinates { (1,50.32)(2,45.41)(3,42.89)(4,41.36)(5,40.38)(6,39.76)(7,39.43)(8,39.33)(9,39.44)(10,39.77)(11,40.39)(12,41.37)(13,42.90)(14,45.43)(15,50.34) };
    
\addplot[
color=red,
mark=x
]
coordinates { (1,42.06)(2,40.56)(3,39.39)(4,38.49)(5,37.82)(6,37.37)(7,37.11)(8,37.04)(9,37.12)(10,37.37)(11,37.82)(12,38.49)(13,39.39)(14,40.57)(15,42.05) };
    
\addplot[
color=black,
mark=x
]
coordinates { (1,48.70)(2,42.51)(3,39.09)(4,36.87)(5,35.27)(6,34.06)(7,33.47)(8,33.19)(9,33.29)(10,33.80)(11,34.88)(12,36.34)(13,38.50)(14,41.75)(15,47.61) };
\legend{Proposed,Linear Interpolation,Bilinear Up-sampling,Nearest-neighbours Up-sampling}
\end{axis}

\end{tikzpicture}
\caption{Angular-PSNR averaged on the LungCT test dataset. Each approach to sinogram enhancement differ. The proposed data-driven interpolation is better than the deterministic ones but the variation from one angular distance to another is not as smooth as the one of deterministic approaches.}
\label{figure:angular_PSNR_LungCT}
\end{figure}

\begin{figure}
\centering
\begin{tikzpicture}
\begin{axis}[
title={Angular PSNR for the LungCT test dataset},
xlabel={Acquisition Index},
ylabel={PSNR [dB]},
xmin=0, xmax=16,
ymin=37, ymax=60,
xtick={1,4,8,12,15},
ytick={42, 55},
legend style={at={(0.5,-0.1)},anchor=north},
ymajorgrids=true,
grid style=dashed,
legend cell align={left},
]
    
\addplot[
color=green,
mark=x
]
coordinates { (1,54.75)(2,50.33)(3,47.88)(4,46.48)(5,45.58)(6,44.20)(7,44.04)(8,43.71)(9,44.00)(10,44.14)(11,45.50)(12,46.45)(13,47.89)(14,50.18)(15,54.80) };
    
\addplot[
color=green,
mark=+
]
coordinates { (1,55.02)(2,50.87)(3,48.62)(4,47.19)(5,46.19)(6,45.39)(7,44.97)(8,44.80)(9,44.94)(10,45.34)(11,46.12)(12,47.12)(13,48.56)(14,50.81)(15,54.98) };
    
\addplot[
color=blue,
mark=+
]
coordinates { (1,52.97)(2,49.32)(3,47.25)(4,46.00)(5,45.20)(6,44.67)(7,44.40)(8,44.32)(9,44.43)(10,44.73)(11,45.26)(12,46.06)(13,47.29)(14,49.31)(15,52.45) };
    
\addplot[
color=red,
mark=+
]
coordinates { (1,48.54)(2,46.70)(3,45.28)(4,44.35)(5,43.69)(6,43.20)(7,42.95)(8,42.88)(9,42.96)(10,43.19)(11,43.66)(12,44.30)(13,45.24)(14,46.67)(15,48.53) };
    
\addplot[
color=black,
mark=+
]
coordinates { (1,50.38)(2,47.77)(3,46.10)(4,45.06)(5,44.31)(6,43.73)(7,43.46)(8,43.34)(9,43.41)(10,43.64)(11,44.17)(12,44.85)(13,45.94)(14,47.61)(15,50.15) };
\legend{Proposed,Improved Proposed,Improved Linear Interpolation,Improved Bilinear Up-sampling,Improved Nearest-neighbours Up-sampling}

\end{axis}
\end{tikzpicture}
\caption{Angular-PSNR averaged on the LungCT test dataset, after 2D enhancement. Our proposed approach is marginally improved by the 2D enhancement procedure, unlike deterministic methods that largely benefit from it.}
\label{figure:angular_PSNR_LungCT_improved}
\end{figure}
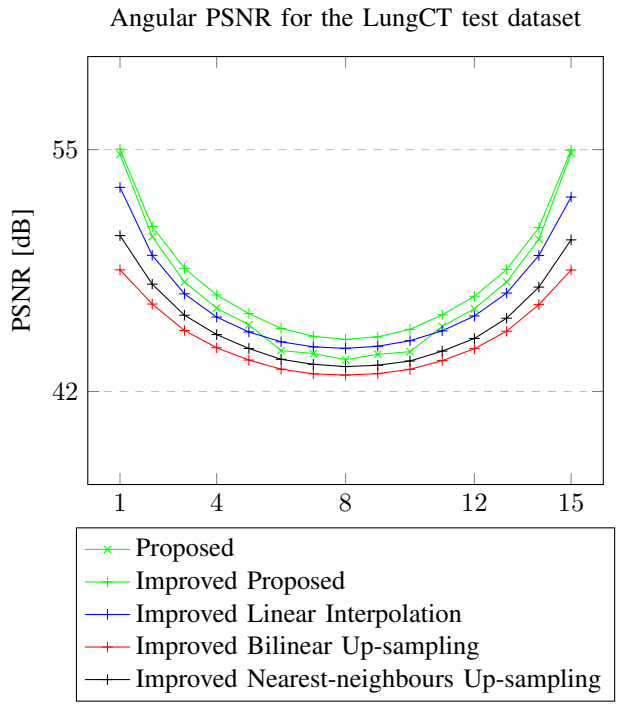

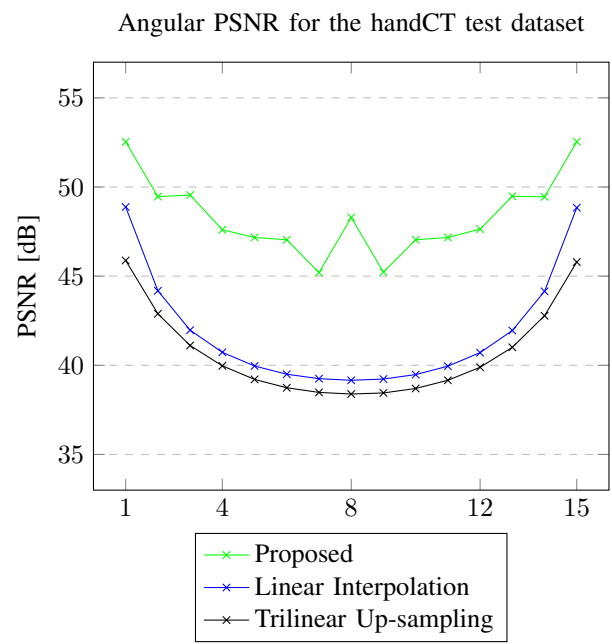
\begin{figure}
\centering
\begin{tikzpicture}
\begin{axis}
[title={Angular PSNR for the handCT test dataset},xlabel={Acquisition Index},ylabel={PSNR [dB]},xmin=0, xmax=16,ymin=33, ymax=57,xtick={1,4,8,12,15},legend style={at={(0.5,-0.1)},anchor=north},ymajorgrids=true,grid style=dashed,legend cell align={left},]
    
\addplot[color=green, mark=x]
coordinates { (1,52.53)(2,49.46)(3,49.55)(4,47.60)(5,47.17)(6,47.04)(7,45.20)(8,48.29)(9,45.23)(10,47.04)(11,47.17)(12,47.64)(13,49.48)(14,49.45)(15,52.54) };
    
\addplot[color=blue, mark=x]
coordinates { (1,48.88)(2,44.18)(3,41.97)(4,40.73)(5,39.96)(6,39.50)(7,39.25)(8,39.16)(9,39.23)(10,39.48)(11,39.95)(12,40.71)(13,41.95)(14,44.15)(15,48.84) };
    
\addplot[color=black, mark=x]
coordinates { (1,45.88)(2,42.89)(3,41.11)(4,39.97)(5,39.21)(6,38.74)(7,38.48)(8,38.39)(9,38.45)(10,38.70)(11,39.16)(12,39.89)(13,41.01)(14,42.78)(15,45.79) };
\legend{Proposed,Linear Interpolation,Trilinear Up-sampling}
\end{axis}

\end{tikzpicture}
\caption{Angular-PSNR averaged on the HandCT test dataset. The proposed approach outperforms the deterministic interpolation procedures for all angular distances considered.}
\label{figure:angular_PSNR_HandCT}
\end{figure}

\begin{figure}
\centering
\begin{tikzpicture}
\begin{axis}
[
title={Sinograms-PSNR against angular interval between acquisitions},
xlabel={Up-sampling Ratio},
ylabel={PSNR [dB]},
legend style={at={(0.5,-0.1)},anchor=north},
xtick={2,4,8,16},
xticklabels={0.7\textdegree , 1.4\textdegree,2.8\textdegree,5.6\textdegree},
ymajorgrids=true,
grid style=dashed,
legend cell align={left},
]
    
\addplot[color=green, mark=x]
coordinates { (2,56.94)(4,54.02)(8,51.61)(16,48.93) };
    
\addplot[color=blue, mark=x]
coordinates { (2,56.99)(4,51.09)(8,46.18)(16,42.44) };
    
\addplot[color=black, mark=x]
coordinates { (2,55.17)(4,48.92)(8,44.38)(16,41.07) };
\legend{Proposed,Linear Interpolation,Trilinear Up-sampling}

\end{axis}

\end{tikzpicture}
\caption{Our approch yields an improvement of the sinogram-PSNR from an up-sampling ratio of 4.}
\label{figure:Sinogram-PSNR_against_ratio}
\end{figure}
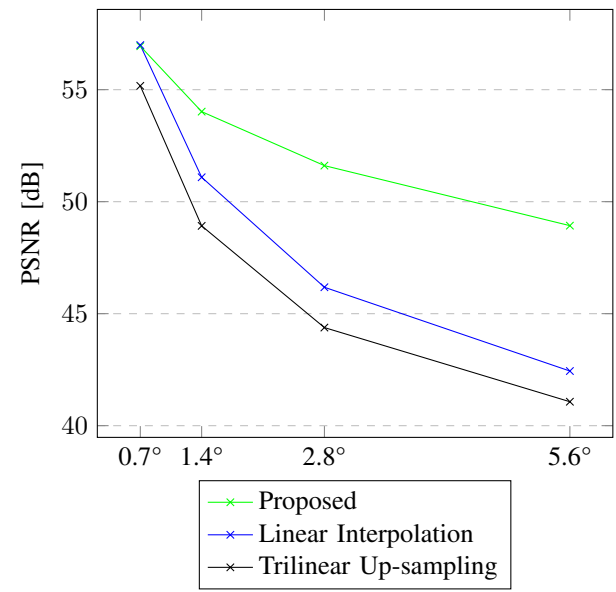

\begin{figure}
\begin{tikzpicture}
\begin{axis}
[
title={Image-PSNR against angular interval between acquisitions},
xlabel={Up-sampling Ratio},
ylabel={PSNR [dB]},
legend style={at={(0.5,-0.1)},anchor=north},
ymajorgrids=true,
grid style=dashed,
xtick={2,4,8,16},
xticklabels={0.7\textdegree , 1.4\textdegree,2.8\textdegree,5.6\textdegree},
legend cell align={left},
]
    
\addplot[color=cyan, mark=x]
coordinates { (2,74.07)(4,64.53)(8,57.55)(16,51.66) };
    
\addplot[color=green, mark=x]
coordinates { (2,81.60)(4,74.17)(8,76.85)(16,72.58) };
    
\addplot[color=blue, mark=x]
coordinates { (2,83.81)(4,77.34)(8,70.92)(16,65.20) };
    
\addplot[color=black, mark=x]
coordinates { (2,80.39)(4,72.85)(8,66.91)(16,62.01) };
\legend{Scarce,Proposed,Linear Interpolation,Trilinear Up-sampling}
\end{axis}

\end{tikzpicture}
\caption{Unlike in Fig. \ref{figure:Sinogram-PSNR_against_ratio}, in the image space, the proposed approach is significantly better than the deterministic interpolation only for an up-sampling ratio of eight or sixteen.}
\label{figure:Image-PSNR_against_ratio}
\end{figure}
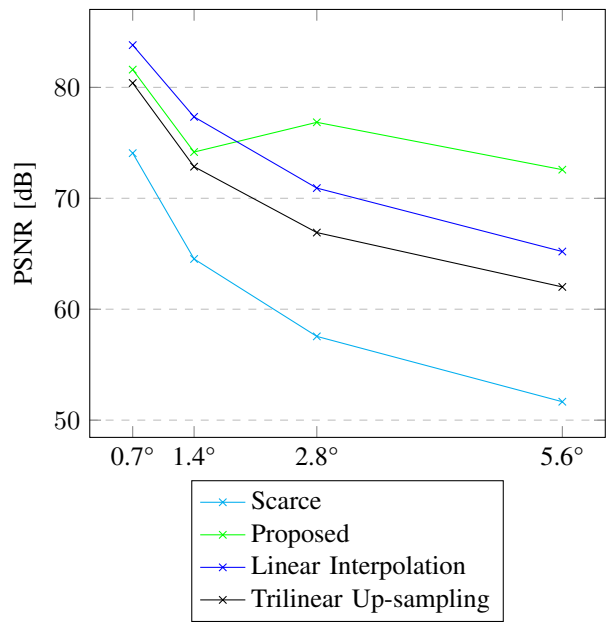